**Multiferroic RMnO₃ thin films**

**J. Fontcuberta**

*Institut de Ciència de Materials de Barcelona (ICMAB-CSIC),
Campus de la UAB, Bellaterra E-08193, Catalonia, Spain*

**Abstract**

Multiferroic materials have received an astonishing attention in the last decades due to expectations that potential coupling between distinct ferroic orders could inspire new applications and new device concepts. As a result, a new knowledge on coupling mechanisms and materials science has dramatically emerged. Multiferroic RMnO₃ perovskites are central to this progress providing a suitable platform to tailor spin-spin and spin-lattice interactions.

With views towards applications, development of thin films of multiferroic materials have also progressed enormously and nowadays thin film manganites are available with properties mimicking those of bulk compounds. Here we review achievements on the growth and characterization of magnetic and ferroelectric properties of hexagonal and orthorhombic RMnO₃ epitaxial thin films, discuss some challenging issues and we suggest some guidelines for future research and developments.

## 1. Introduction

Multiferroics materials have had an extraordinary renaissance in the last decades. Some few experimental results broke the relatively peaceful state of multiferroics research by drawing the attention of the whole scientific community interested on functional oxides. The report of a large polarization in antiferromagnetic $BiFeO_3$ perovskite thin films [1] stimulate a blossoming activity. Although later experiments on single crystals [2] indicated that the large polarization was not due to epitaxial strain as initially suggested, the race towards multiferroics thin films in the search of large magnetoelectric coupling had started. In fact, in $BiFeO_3$ the Curie temperature where ferroelectric sets in (~1103 K) and the Néel temperature where long range antiferromagnetic order develops (643 K), occur at temperatures differing by about 460 K, indicating that different mechanism triggers both orders and anticipating a weak (magnetoelectric) coupling, among them. Hexagonal manganites $RMnO_3$ were also known to be robust room-temperature ferroelectrics and developing an antiferromagnetic ordering at low temperature (typically around 70-100 K) (see Lorentz et al. [3] for a recent review) and, similarly to the $BiFeO_3$ case, weak magnetoelectric coupling is expected.

Hybrid structures containing magnetic (ferromagnetic or antiferromagnetic) and ferroelectric materials, either in the form of nanocomposites or heterostructures were early shown to be a promising way to circumvent the intrinsically weak magnetoelectric coupling in $BiFeO_3$ [4] and in hexagonal manganites [5-7], and it was shown that they could be useful for electric manipulation of ferromagnetic layers in magnetic devices. The activity on these hybrid structures has led to a number of impressive results that may have a long term impact on spintronics [8]. Progress has been revised recently and we drive the reader to appropriate overviews [9-13].

Almost simultaneously with the $BiFeO_3$ rediscovery, Kimura et al. [14] reported that in $TbMnO_3$ perovskite, a polarization develops related to the particular magnetic ordering existing in this oxide; consequently, both order parameters can be manipulated by external magnetic fields, giving rise to a giant magnetoelectric coupling. This finding constituted a hallmark in solid state physics and materials physics as it indicated the existence of a so-far unknown mechanism for electric polarization arising from magnetic order and suggested new ways for searching materials with a large magnetoelectric coupling. Unfortunately, the magnetic order in TbMnO$_3$ and other similar Mn-based perovskites occurs at low temperatures (typically below 40 K) and thus practical applications appear to be difficult. However, these oxides offered a window to explore and discover new mechanisms of ferroelectricity and magnetoelectric coupling. The progress has been impressive and multiferroic properties of bulk Mn-based perovskites and related materials have been recently reviewed elsewhere [15, 16]. BiMnO$_3$ and Bi$_2$MM'O$_6$ constitute a particular case of multiferroic Mn-based material: both the origin of its ferroelectric character (the lone-pair or Bi$^{3+}$) and the nature of the magnetic ordering (ferromagnetic) are radically different that those of non-Bi containing RMnO$_3$ oxides. Although thin films of these compounds [17, 18] have allowed to obtain devices with new functionalities [19], they will not be revised here. Bi-based manganites have been reviewed recently by A. A. Belik [20].

An overview of preparation of multiferroic thin films has been released by Martin et al [21], and Lawes and Srinivasan [22] have prepared an excellent overview of their properties. Here we will concentrate on properties of RMnO$_3$ thin films. Mn-based perovskite have an orthorhombic (o-RMnO$_3$) or hexagonal (h-RMnO$_3$) unit cell symmetry depending on the size of the R-ion (R is a lanthanide or Y) and constitute by itself a universe whose magnetic and dielectric properties, when prepared in thin film form, may largely differ from those encountered in bulk and are strongly dependent of fine details of thin film texture, strain state or thickness. This sensitivity leads to some important questions that demand dedicated attention. For instance, how size effects or strain in thin films may affect the cycloidal antiferromagnetic order, which is known to be essential for the occurrence of ferroelectricity in some orthorhombic perovskites? Or, can the delicate competition of steric effects leading to hexagonal or orthorhombic RMnO$_3$ structures and the inherently different magnetic interactions and their competition, be modified by epitaxial strain?

The review is structured as follows. We first briefly summarize the structure and magnetic (Sections 2) and ferroelectric properties (Section 3) of bulk RMnO$_3$ oxides. In Section 4 we overview the growth and ferroelectric and magnetic properties of hexagonal h-RMnO$_3$ films (Section 4.1) and the growth of orthorhombic o-RMnO$_3$ films (Section 4.2), with special emphasis on film's textures and strain states. Section 5 is devoted to the multiferroic properties of o-RMnO$_3$ thin films and we end (Section 6) with a brief summary and a discussion on some open questions deserving further consideration.

## 2. Structural and magnetic properties of Mn-based perovskites and related structures
### 2.1. Structural considerations

The crystal structure of RMO$_3$ perovskite oxides , where R is typically a alkaline–earth or a lanthanide ion and M a transition metal, can be viewed as deriving from a compact arrangement of a simple cubic corner-sharing MO$_6$ octahedra forming a cage with the R ion occupying its center. The M-O-M bond angle and to lesser extent, the M-O bond length, are determined by the size of the R-ions. The reason is that the R ions can be smaller than the available space at its cage and, as a result, the O$^{2-}$ network may distort by a cooperative rotation and buckling of the MO$_6$ octahedra to reduce the volume available to the R ion. For a perfect cubic perovskite the M-O-M bond angle is 180°. The relative lengths of the R-O and M-O bonds determine the stability of the cubic structure. The Goldschmidt tolerance factor $t = d$(R-O)$/d$(M-O)$\sqrt{2}$, where $d$(R-O) and $d$(M-O) are the actual R-O and M-O bond lengths (that can be conveniently estimated by $d$(R-O)= r(R)+ r(O) and $d$(M-O) = r(M)+ r(O), being r(R), r(M) and r(O) the corresponding ionic radii), is a convenient measure of the compactness of the structure and thus of its stability. For a cubic perovskite $t$ =1 whereas for $t$ <1 the symmetry is reduced and the M-O-M bond closes away from 180°. This occurs when the R-size is reduced and the R-O bond length shrinks. The MO$_6$ octahedra rotate to reduce the available space at the central cage for the R-ion and the symmetry becomes orthorhombic (space group Pbnm). This distortion is known as GdFeO$_3$-type distortion. For smaller R-ions, the symmetry reduced to rhombohedral. For even smaller R-ions, the connectivity of the MO$_6$ network cannot be further preserved and the symmetry is reduced to hexagonal and the structure is not longer a perovskite but formed by layers of $MO_6$ triangular bipyramids. In **Fig. 1** we list the stable structures of $RMnO_3$ oxides, at ambient conditions, as a function of the R-ion size. We strength that for $r < r(Dy^{3+})$, that is for R= Ho, Er, Tm, Yb, Lu and Y. the stable form of these $RMnO_3$ oxides is hexagonal. However it was shown long ago, that high pressure synthesis [23], soft-chemistry [24] or epitaxial strain in thin films [25] allowed to obtain metastable orthorhombic forms of all these oxides.

| Stable bulk **orthorhombic** | | | | | | | Stable bulk **Hexagonal** | | | | | |
|---|---|---|---|---|---|---|---|---|---|---|---|---|
| LaMnO₃ | NdMnO₃ | SmMnO₃ | EuMnO₃ | GdMnO₃ | TbMnO₃ | DyMnO₃ | HoMnO₃ | ErMnO₃ | TmMnO₃ | YbMnO₃ | LuMnO₃ | YMnO₃ |

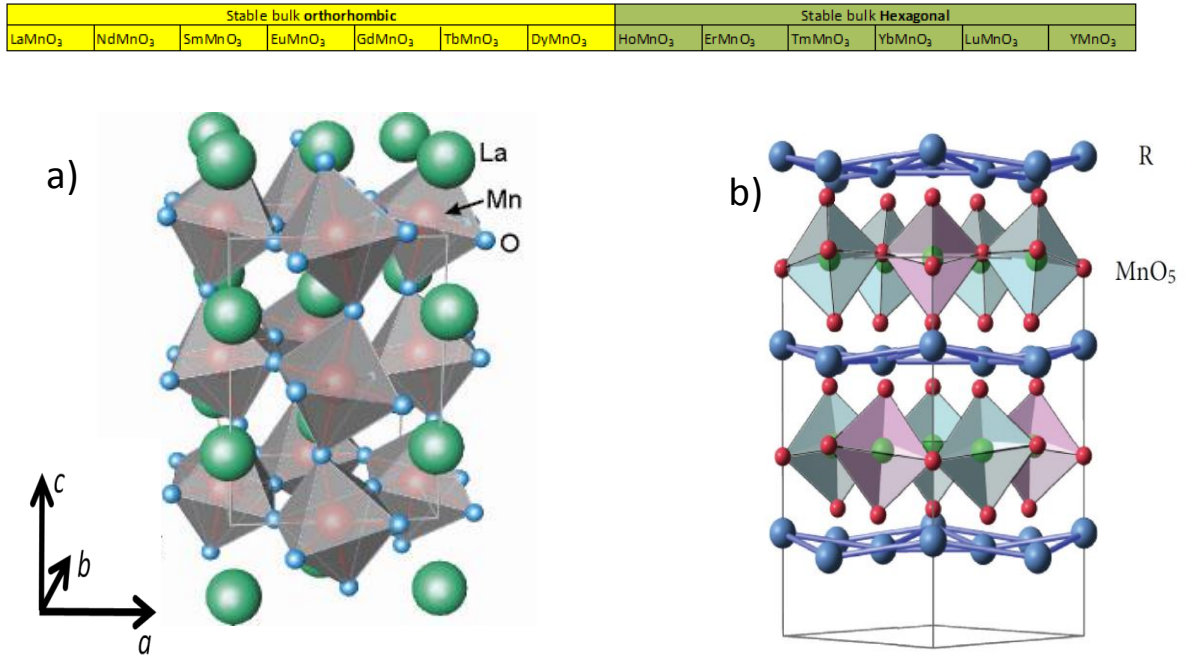

**Fig. 1.** *Structures of (a) orthorhombic LaMnO₃ and (b) hexagonal RMnO₃ manganites (adapted from [26] and [3])*

In addition to rigid rotations and buckling of $MO_6$ octahedra, in case the M-ions are Jahn-Teller active, such as high-spin $Mn^{3+}$ ($d^4$: $t_{2g}^3 e_g^1$), the $MO_6$ octahedra can also distort to lift the degeneracy of half-occupied $e_g^1$ electronic orbitals. This implies creating shorter and longer M-O bonds within the $MO_6$ octahedra. At lower temperature this bond-pattern freezes in an orbitally ordered state $3x^2$-$r^2$/$3y^2$-$r^2$ leading to distinct cell parameters, although without reducing the *Pbnm* symmetry further. These orthorhombic structures are known as O and O' respectively. The Jahn-Teller distorted phase (O') thus results from cooperative distortions of $MO_6$ octahedra superimposed to their rigid rotation and buckling. Whereas in the GdFeO₃-like distorted phase (O) $c/\sqrt{2} > a$ and $a < b$ ( ($a,b,c$) are the cell parameters), in the O' phase $c/\sqrt{2} < a < b$. The temperature at which the JT-related distortion blocks depends on the R-size, whereas its amplitude, which is largely dictated by electrostatic interactions within the $MO_6$ octahedra, does not vary significantly with R-size.

In the paraelectric phase of hexagonal $RMnO_3$ the trigonal symmetry of the $MnO_5$ coordination polyhedron breaks the degeneracy of the 3d levels by forming three sets of states: an $e_{2g}$ doublet composed of $d_{xz}$ and $d_{yz}$ states, an $e_{1g}$ doublet formed by $d_{xy}$ and $d_{x2-y2}$ states and a singlet $a_{1g}$ state formed by $d_{z2}$ states. As in the Mn-O distance is shorter along c-axis than in the basal plane, the $d_{z2}$ are the highest in energy. Therefore, electronic configuration should be: $e_{2g}^2 e_{1g}^2 a_{1g}^0$ ; as the uppermost occupied levels of Mn-3d⁴ are fully occupied these ions are not Jahn-Teller active and the symmetry of the MnO$_5$ is not further reduced.

## 2.2. Magnetism and magnetic interactions

In perovskites where M is a 3d metal, the 3d-orbitals are somehow hybridized with neighboring 2p-O orbitals, forming bonding and antibonding states. It turns out that the fully occupied bonding states are primarily of 2p-O parentage and the antibonding states are mostly 3d-like and contain the spins that are responsible for magnetic interactions. As a result of the electrostatic interaction of 3d orbitals with neighboring O$^{-2}$ ions, their degeneracy is partially lifted by the cubic crystalline field leading to a triplet (t$_{2g}$: d$_{xy}$,d$_{xz}$,d$_{yz}$) state and a doublet (e$_g$: x$^2$-y$^2$; 3z$^2$-r$^2$) state. Subsequent reduction of symmetry by JT-like distortion would further split the degeneracy of these triplet and doublet states as described above. In 3d-metal oxides the spin-orbit coupling is weak and the orbital moment is virtually quenched. Consequently, the magnetic moment of an M ion is simply given by 2μ$_B$S where S is the total spin as dictated by the Hund rules, μ$_B$ is the Bohr magneton and we have assumed that the Landé factor is 2.

In 3d-RMO$_3$ perovskites, two types of magnetic interactions are relevant. The first one, so-called *superexchange interaction*, is the typical in systems where 3d orbitals form fully occupied bands and are thought as virtual charge-transfer between M-O-M ions. Those are the relevant interactions in magnetic oxide insulators. The second, named *double-exchange*, involve partially occupied 3d-orbitals and thus is characteristic of metallic magnetic oxides.

Taking 3d$^4$-Mn$^{3+}$ as example, the electronic configuration in a cubic crystal field is 3d-t$_{2g}^3$e$_g^1$ and the degenerated e$_g$ doublet (x$^2$-y$^2$ and 3z$^2$-r$^2$) is further broken by the JT interaction. The electronic state of the lowest energy, for instance x$^2$-y$^2$, forms a sub-band fully occupied, thus anticipating an insulating behavior, with its spin parallel to the core (t$_{2g}^3$) spins (Hund rule). Therefore the magnetic interactions in insulating 3d$^4$-RMnO$_3$ oxides are described in terms of superexchange interactions and they determine the magnetic order. In an ideal cubic perovskite, the strength of the direct interaction between 3d-metallic ions M-M is weaker than that of indirect interactions (J$_{nn}$) between metal nearest-neighbors (nn) mediated by oxygen ions, i.e. M-O-M bonds along bridging corner-sharing octahedra. Similarly, the strength of the nn M-O-M interactions is larger than that of the next-nn (nnn) along an orthogonal direction. The strength and sign of the magnetic M-O-M interactions are strongly dependent on the number of electrons of the 3d-orbitals involved (3d$^n$) and on the M-O-M bonding angle (θ) and involve the oxygen 2p orbitals. The sign and strength of J$_{nn}$ are dictated by Goodenough-Kanamori rules (G-K). In RMnO$_3$ perovskites with relatively large R-ion size and thus small GdFeO$_3$-like distortion, such as LaMnO$_3$, the cooperative JT distortion produces a staggered 3d(3x$^2$-r$^2$)-2p-3d(3y$^2$-r$^2$) orbital pattern in the *ab* plane. According to the G-K rules, a ferromagnetic (FM) interaction 3d(3x$^2$-r$^2$)-2p-3d(3y$^2$-r$^2$) sets in between these e$_g^1$ spins (J$^{FM}_{nn}$(e$_g$) > 0) and a weaker antiferromagnetic (AF) interaction between the t$_{2g}^3$ spins (J$^{AF}_{nn}$(t$_{2g}$) < 0). The latter is dominant along the *c*-axis. Therefore, spins in the basal *ab* plane are coupled ferromagnetically but antiferromagnetically along the *c*-axis (**Fig. 2a**). This is the so-called A-type AF ordering, which is characteristic of RMnO$_3$ perovskites with small Mn-O-Mn bond bending and appearing at T$_N$ around 40 K, depending on the lanthanide ionic radius. When θ gradually bends by reducing the R-size, J$^{FM}_{nn}$(e$_g$) is also reduced, and importantly, antiferromagnetic next-neighbor interactions (M-O-O-M) J$^{AM}_{nnn}$(e$_g$) < 0 become gradually more important, thus competing with the in-plane ferromagnetic J$^{FM}_{nn}$(e$_g$) ( > 0) interactions and introducing magnetic frustration. For the most distorted Mn-O-Mn bonds, such as HoMnO$_3$, this give rise to an up-up-down-down collinear spin order in the *ab*-plane (E-type AF order), setting in at T$_N$ around 40 K, depending on the lanthanide ionic radius (**Fig. 2b**). Not surprising, between these two limiting cases, the competition of magnetic interactions lead to more complex magnetic orders, such as cycloidal order, in which the spins can rotate either within the *ab*-plane (*ab*-cycloids) or within the *bc* plane (*bc*-cycloids) both propagating along the *b*-axis or collinear with sinusoidal modulation (**Fig. 3b,c,d**). As shown in the **Fig. 3a**, the magnetic ordering temperatures and the nature of the magnetic order itself, are determined by the Mn-O-Mn bond angle and correspondingly by the $R^{3+}$ ionic radius. At still lower temperature, the contribution of $R^{3+}$ magnetic moments may modify the ordering of the $Mn^{3+}$ ions.

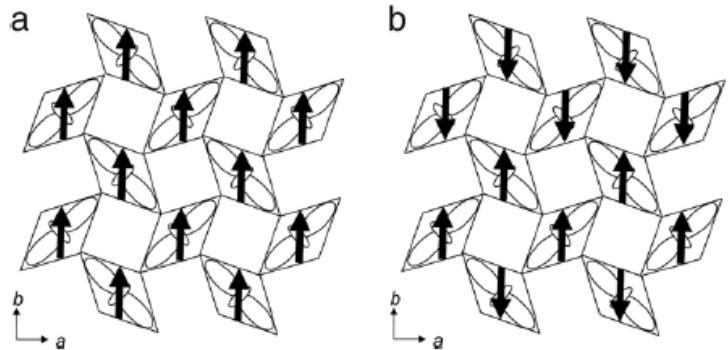

*Fig. 2 a) Antiferromagnetic A-type ordering. Spins are ferromagnetically coupled within the ab-plane and antiferromagnetically coupled along the c-axis. b) Antiferromagnetic E-type order. The spins are antiferromagnetically coupled along the b-axis and c-axis but ferromagnetically coupled along a-axis forming up-up-down-down chains along [110] (adapted from [27]*)

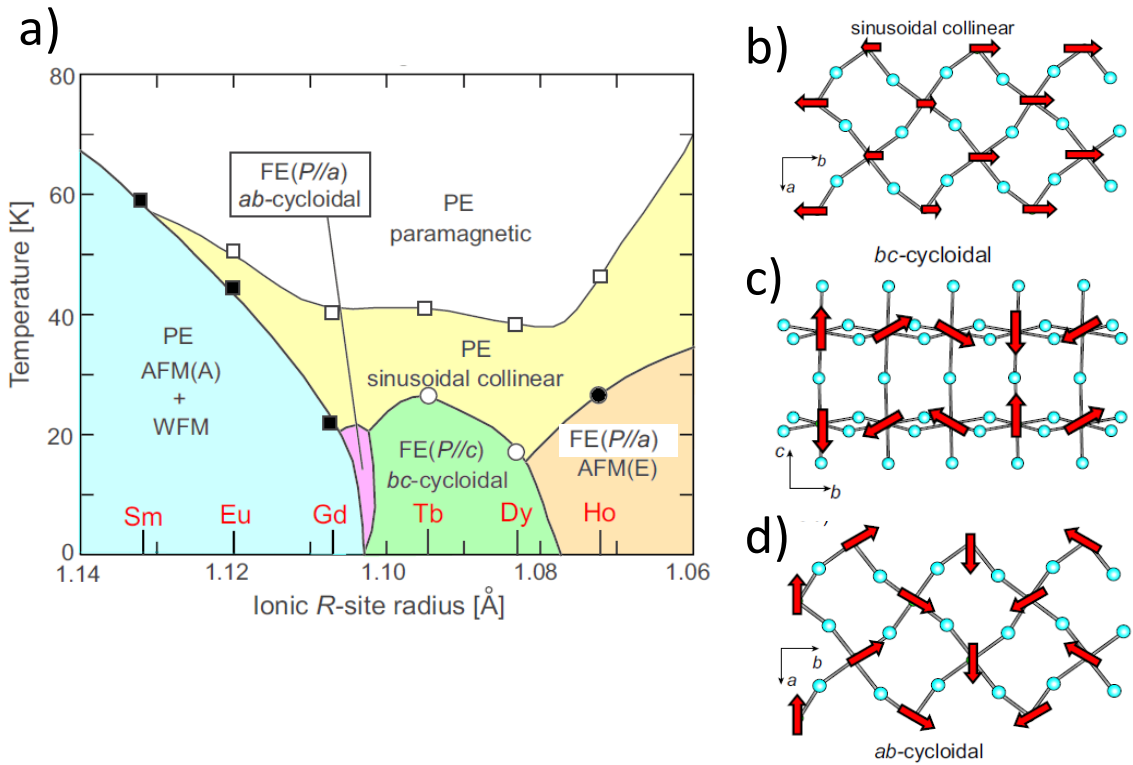

*Fig. 3 a) Magnetic phase diagram of RMnO₃ perovskites as a function of the ionic size of the $R^{3+}$ ion. Open and closed squares indicate the Néel temperature and the temperature where the wavelength of the modulated magnetic structure blocks (lock-in) respectively. The direction of the polarization is indicated (P). Panels b), c) and d) illustrate: sinusoidal collinear spin arrangement (E-type), bc-cycloidal and ab-cycloidal spin order respectively. Adapted from [28].*

As mentioned above, for manganites with even smaller lanthanides, (R= Y, Ho, Er, Tm, Yb, Lu) the stable structure is hexagonal ($P6_3mmc$) where the $Mn^{3+}$ ions occupy the center of a triangular bipyramid MnO₅, that shares the corners of the trigonal basal plane O atoms thus forming a hexagonal bidimensional $MnO_3$ array. This paralelectric phase transforms into the non-centrosymmetric polar group ($P6_3cm$) at about 1300 K. The Mn spins are confined to the bidimensional $MnO_3$ plane and order antiferromagnetically at the Néel temperature of about ≈ 60 K- 120 K (depending on the lanthanide) forming a triangular spin lattice with a magnetic symmetry that depends on R and temperature [3, 29] (Fig. 4). In compounds with magnetic rare earth elements, a spin reordering transition occurs at much low temperature $T_{RE}$ (4 − 8 K) which produce a further change of magnetic symmetry. In the particular case of $HoMnO_3$, the magnetic phase diagram is more complex as an additional spin reorientation, involving in-plane rotation of Mn-spins takes place at about around 40 K.

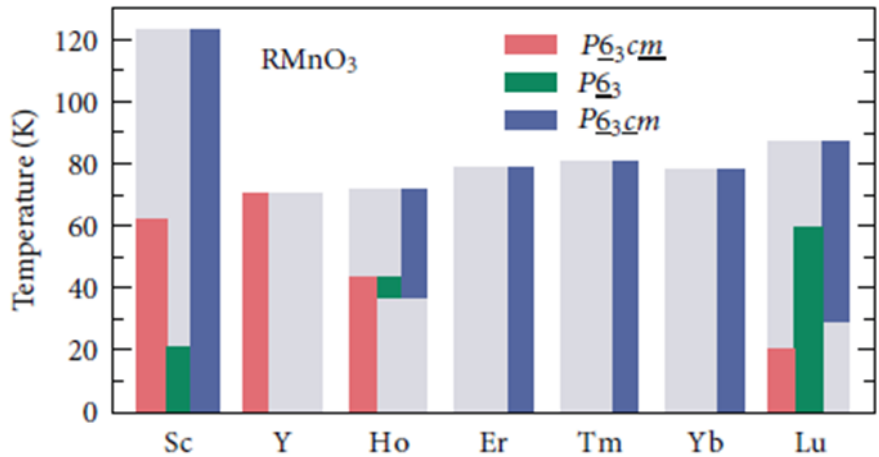

**Fig. 4** *Néel temperatures and magnetic symmetries of hexagonal $RMnO_3$ oxides. Adapted from [29]*

As mentioned, either in the orthorhombic or in the hexagonal phases, the sublattice containing magnetic rare-earth, also orders magnetically at much lower temperature (around 10 K or below). Still, although the rare-earth sublattice is paramagnetic-like at higher temperature, its contribution to the high-temperature (> $T_N$) paramagnetic susceptibility of the $RMnO_3$ oxides can be very high due to the large magnetic moment of rare-earths and its magnetic anisotropy.

### 3. Ferroelectricity in Mn-based perovskites and related structures

Whereas magnetic ordering described above breaks time-reversal symmetry, ferroelectricity requires loosing the symmetry center. Hexagonal manganites (h-$RMnO_3$) belong to the so called type-I multiferroics, where the ferroelectricity and magnetism have different origins and set it at largely different temperatures. Orthorhombic perovskites (o-$RMnO_3$) belong to type-II multiferroics where ferroelectricity and the concomitant non-centrosymmetric character result from the magnetic order. In type II multiferroics, the centre of symmetry of positive and negative charges does not coincide due to: a) non-collinear magnetic structures and asymmetric exchange, b) symmetric exchange and striction or c) spin-dependent hybridization. Of major interest for o-$RMnO_3$ ferroelectric perovskites are mechanisms a) and b). These mechanisms have been extensively reviewed elsewhere in this issue (see also [30, 31] and [16]) and documented in detail by exhaustive experimental works on single crystals and polycrystalline samples. Here we will restrict to summarize the main ideas.

In a magnetically ordered structure and in presence of a significant spin-orbit coupling, it can be expected that the position of anions bonding neighboring metal ions can be shifted away from the equilibrium position when tilting the atomic spins. If this shift occurs in-phase across the lattice, it may lead to a net shift of the center of mass of positive and negative charges thus producing a macroscopic dipole (the polarization, **P**), contained in the plane where spins rotate, which can be written as: $P \approx A \sum r_{ij} \times (S_i \times S_j)$ where $S_i$, $S_j$ are the spins of neighboring ions, $r_{ij}$ the vector connecting them and A is a spin-orbit related parameter. It follows that the amplitude of the induced polarization will depend on the strength of the spin-orbit coupling and the relative angle of neighboring spins. This is what happens in the cycloidal magnetic structures, where a polarization appears in the cycloidal plane and perpendicular to the propagation direction of the cycloid (see Fig. 5a).

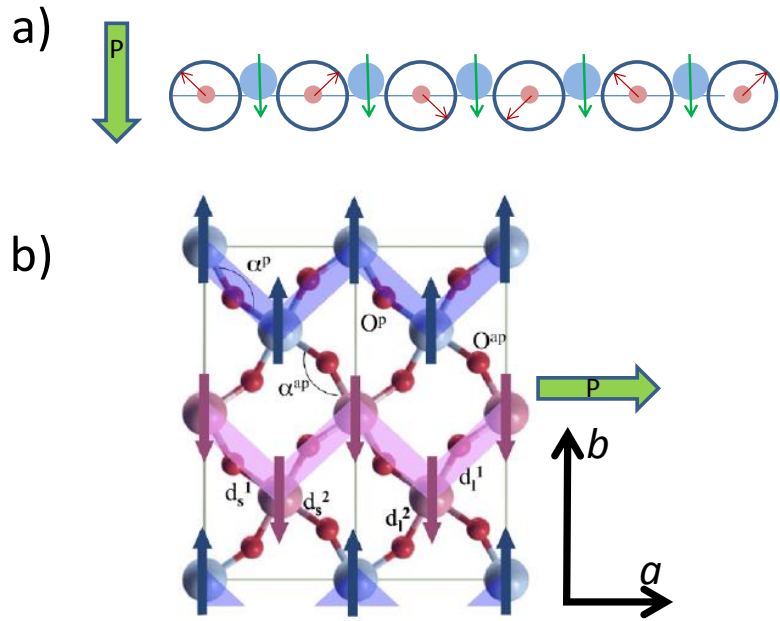

**Fig. 5** *a) Cycloidal order developing a finite dipole in each bond that add in-phase leading to a finite polarization perpendicular to the propagation direction of the cycloid and contained in its rotation plane. b) Collinear spin ordering in E-type antiferromagnet creating a finite polarization along a-axis. (Adapted from* [32])

As the cycloidal magnetic order results from a delicate competition of FM and AFM interactions, it shall be expected that it can be modified by a magnetic field and subsequently the polarization can be extremely sensitive to magnetic fields. The observation of polarization and magnetic-field induced flopping of **P**, first reported in TbMnO$_3$ single crystals by Kimura et al. [14], constituted a hallmark in the field. Naturally, this mechanism should not operate in case of collinear spins. This is the case of A-type AF, such as LaMnO$_3$, NdMnO$_3$ and SmMnO$_3$ (in decreasing order of ionic radius of the lanthanide) where no ferroelectricity is observed (Fig, 3). More interesting is the case of smaller R ions, such as HoMnO$_3$, which as described above can be stabilized in the o-RMnO$_3$ phase, which is a collinear E-type antiferromagnet. Here, it has been shown that ferroelectricity results from the distinct striction appearing between the up-up (or down-down) and up-down (or down-up) spin pairs along the spin up-up-down-down chain [32-34] (Fig. 4b). In HoMnO$_3$ ferromagnetically coupled spins chains are along the *a*-axis and antiferromagnetic chains are along the *b*-axis, and thus up-up-down-down chains align along the diagonal of the *ab-plane* and the polarization appears along the *a*-axis. Here, the strong exchange striction which results from exchange interactions responsible of the collinear magnetic ordering anticipates a larger polarization than in cycloidal manganites but a weaker effect on magnetic fields on polarization. Indeed it is found that in bulk materials, the polarization of cycloidal ferroelectric perovskites is of about 0.2 μC/cm$^2$ and can be drastically modified by a magnetic field and in E-type ferroelectrics it is of about 0.5 µC/cm² and less sensitive to magnetic fields.

The magnetic ordering in antiferromagnetic perovskites comes from the competition of magnetic interactions of different sign in a single magnetic sublattice ($MnO_6$ network) and thus order is established at relatively low temperatures (around 40 K). Ferroelectricity subsequently develops at temperatures well below 40 K. We note that complex not collinear structures in oxides can be obtained in more complex crystallographic structures, when magnetic competition arise from the existence of distinct magnetic sublattices. This is what occurs in magnetic ferroelectrics based on spinel (i.e. $CoCr_2O_4$ [16] or hexaferrite-based oxides [35] or some uncommon iron-oxide phase [36, 37]. Research on thin films of these materials is more seldom attempted and will not be reviewed here.

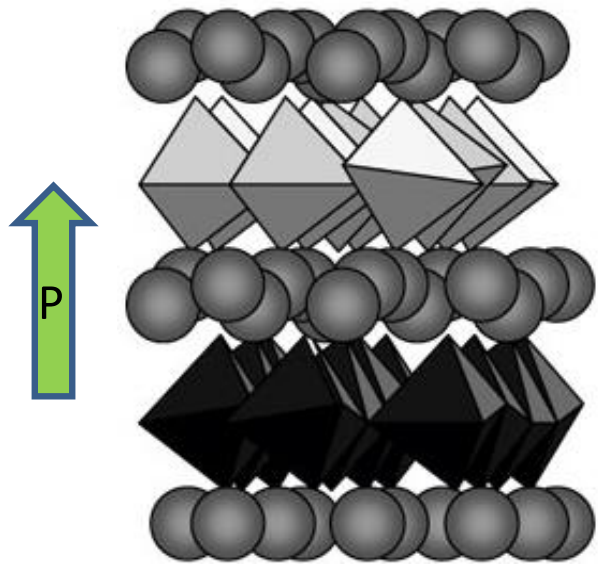

**Fig. 6** *The crystal structure of $YMnO_3$ in the ferroelectric phases. The trigonal bipyramids depict $MnO_5$ polyhedra and the spheres represent Y ions. (Adapted from* [38])

In contrast to the orthorhombic manganites described above, ferroelectricity in hexagonal manganites develops at much higher temperature than magnetic order. Indeed, ferroelectricity sets in typically over 900 K whereas antiferromagnetic order develops around 70 K - 130 K. In hexagonal manganites ferroelectricity appears from pure steric effect, related to the tilting of rigid $MnO_5$ bipyramids that produces a shift of oxygen ions out of the *ab*-plane that suppresses a mirror plane perpendicular to c-axis, reducing the symmetry from $P6_3/mmc$ to $P6_3cm$ [38] and promoting a polarization of around 5 µC/cm², i.e. about a factor 10 larger than in the magnetic ferroelectrics described above. Due to the different origin of ferroelectricity and magnetic order and the different energy scales associated, the coupling of both order parameters is rather weak.

## 4. Growth of $RMnO_3$ thin films

$RMnO_3$ materials can be roughly categorized as orthorhombic (o-$RMnO_3$) or hexagonal (h-$RMnO_3$). It has been found that epitaxial films of each phase can be obtained by selection of appropriate single crystalline substrates. Moreover, as it will be shown latter, the epitaxial strain induced by substrates can be used to obtain metastable phases of some of these oxides. For instance, in thermodynamically equilibrium, the $RMnO_3$ oxides with small R-ions, such as $YMnO_3$ or $ScMnO_3$ and $RMnO_3$ (R= Ho, Er,Tm, Yb, Lu, Y) are hexagonal. However, the corresponding o-RMnO$_3$ phases have also been obtained by exploiting epitaxial strain [25, 39]. Similarly, RMnO$_3$ with large R-ions (R= La, Nd, Sm, Eu, Gd, Tb, and Dy) are orthorhombic, and some of them, lying closer to the verge of stability of the o-RMnO$_3$/h-RMnO$_3$ phase, such as DyMnO$_3$, TbMnO$_3$ or GdMnO$_3$, have also been stabilized in thin film as h-RMnO$_3$ phase [40-42].

Moreover, suitable selection of the substrate also allows selection of the texture of the thin film, that is, the direction in which the films grows. For instance, o-RMnO$_3$ films with the *a, b,* or *c* axes perpendicular to the film surface, have been obtained. This possibility is of relevance when attempting to discriminate the microscopic origin of ferroelectricity. In the following we will describe the rational of these approaches

### 4.1. Growth of h-RMnO$_3$ thin films and their ferroelectric and magnetic properties

Bulk RMnO$_3$ with small R-ions (R = Ho, Er, Tm, Yb, Lu, Y) are thermodynamically stable in the hexagonal form. Therefore, growth of epitaxial thin films of these materials has been achieved long ago. On the contrary, the stabilization the hexagonal form of orthorhombic bulk manganites RMnO$_3$ with large R-ions (R = La, Nd, Sm, Eu, Gd, Tb, Dy) is more challenging. We first review (i) the growth and characterization of h-RMnO$_3$ films whose parent bulk compound is also hexagonal and next (ii) the epitaxially h-RMnO$_3$ films whose parent bulk compound is orthorhombic.

#### 4.1.1.   h-RMnO$_3$  thin films of hexagonal parent compounds (R = Y, Er, Ho)

The growth of thin films of h-YMnO$_3$ oxides has been recently reviewed [21]. h-YMnO$_3$ films were first grown by rf-sputtering by Fujimura et al. [43] and subsequently, films have also been produced by spin-coating, sol-gel, chemical vapor deposition, molecular beam epitaxy and pulsed laser deposition. Used substrates include MgO(111), ZnO(0001), Al$_2$O$_3$(0001), (111)YSZ, Pt/TiO$_x$/SiO$_x$/Si, Pt/ZrO$_2$/TiO$_2$/Si and (111)STO. It turned out that films are found to be ferroelectric although with a reduced polarization compared to bulk (1.7 µC/cm$^2$ [44] *vs* 5.5 µC/cm$^2$ [45], respectively). Temperature-dependent magnetization measurements on h-YMnO$_3$ thin films are challenged by the small signal expected from an antiferromagnetic film, further reduced by the bidimensional triangular arrangement of Mn$^{3+}$ spins, and the presence of parasitic signals from substrates that may blur the features associated to the antiferromagnetic transition.  Wu et al. [46] reported magnetic susceptibility data of rather thick films (≈ 180 nm) and observed features occurring at temperature well below the Néel temperature observed in single crystals (≈70 K) and an unexpected hysteretic field-cooled (FC) and zero-field cooled (ZFC) magnetization *vs* temperature data. This hysteresis indicates the presence of some ferromagnetic response in the film as it will discussed later. Neutron diffraction has been used to monitor the onset of long magnetic order in rather thick films (≈ 450-500 nm)  (**Fig. 7a**)   [47]. h-YMnO$_3$ epitaxial films have also been grown on Pt-coated (111)STO [48]. Although some data are available [49], chemical substitutions at either A-site or B-site in thin films have not been systematically studied.

Other hexagonal films: h-ErMnO$_3$ [47, 50] and h-HoMnO$_3$ [47, 51, 52] have also been grown using Pt/(0001)Al$_2$O$_3$ and/or (111)YSZ as substrates. In h-HoMnO$_3$ the polarization loops P(E), which could be measured only at temperatures below 220 K due to leakage, gave a remnant polarization (P$_r$)  of about 3.7 µC/cm$^2$ at 40 K, which is significantly lower that the room-temperature reported value (5.6 µC/cm$^2$). Upon warming, the P(E) loops developed a antiferroelectric-like features (**Fig. 7b**), attributed to domain-pinning by defect dipoles [53]. In h-ErMnO$_3$, the remnant polarization was found to be 3.1 µC/cm$^2$ at 140 K [50]. Similarly to the case of h-YMnO$_3$ above, these values of polarization are smaller than values reported for the corresponding bulk hexagonal materials. It has been reported that in h-RMnO$_3$ epitaxial films (R =Y, Er, Ho and Dy) [54] nanodomains of crystallites with the corresponding hexagonal and polar axis not aligned along the film normal exist. These nanodomains were identified in films of different composition and grown by different techniques (MOCVD and PLD) thus suggesting that their presence in h-RMnO$_3$ films could be rather ubiquitous. This could provide an explanation for the commonly observed reduction of polarization [54].

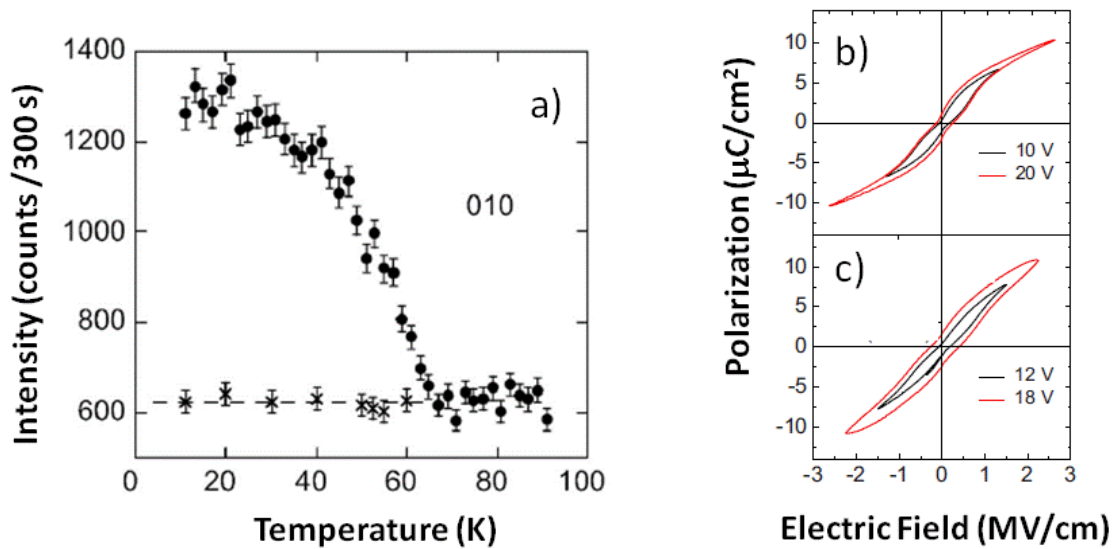

**Fig. 7** (a) Temperature dependence of the (010) Bragg peak of h-YMnO$_3$ thin film. Adapted from [47]. (b,c) P-E loops measured at 70 K for h-HoMnO$_3$ films on (b) Pt/Al2O3 and (c) Pt/TiO2/SiO2/Si substrates. Adapted from [53].

Neutron diffraction was used to confirm the antiferromagnetic character of h-RMnO$_3$ (R = Y, Ho, Er) films [47]. From this study it turned out that in spite that the used films were rather thick (≈ 500 nm) the observed Néel temperatures do not match those of the corresponding bulk materials but are lower (for h-ErMnO$_3$ by a -15 %). Moreover, in the case of h-HoMnO$_3$, the temperature-dependent magnetic response does not reveal the succession of spin ordering and reorientation temperatures observed in bulk. In h-RMnO$_3$ films, systematic studies of the Néel temperature and other fine details of the magnetic ordering are not yet available.

Magnetization measurements on both h-HoMnO$_3$ and h-ErMnO$_3$ films, as much as in the h-YMnO$_3$ films described above, evidenced the presence of an hysteretic (T < 45 K) field-cooled (FC) and zero-field cooled (ZFC) magnetization *vs* temperature response [50], [46]). This unexpected behavior was found to be most noticeable in samples prepared at low O$_2$ (< 200 mTorr) pressure and it was attributed to oxygen vacancies and subsequent magnetic disorder in the Mn-O sublattice (Jang 2008).

Summarizing, although the ferroelectric character of the h-RMnO$_3$ films (R =Y, Ho, Er) is well established, the reported P$_r$ values are below bulk ones. To what extent this reduction is due to substrate-induced strain or other effect is not known. Similarly, the observation of antiferromagnetic-like P(E) loops, although theoretically predicted [38, 55] deserves more attention. The effects of epitaxial strain on magnetic ordering (or reordering) temperatures are largely unknown. Ferromagnetic-like features in the magnetization response in thin films is rather common whose origin is not well settled.

### 4.1.2. h-RMnO$_3$ thin films of orthorhombic parent compounds (R = Sm, Eu, Gd, Dy, Tb)

Hexagonal-phase thin films of RMnO$_3$ whose bulk equilibrium phase is orthorhombic has also been obtained by epitaxial stabilization: (R = Sm, Eu, Gd, Dy [56, 57] and more recently h-TbMnO$_3$ and h-DyMnO$_3$ have also been reported [41]. It has been shown that h-TbMnO$_3$ and

h-DyMnO$_3$ are ferroelectrics with a polarization of about 1.6 μC/cm$^2$. This value cannot be compared to a corresponding bulk value as TbMnO$_3$ is not stable in hexagonal form. However it is comparable to polarization obtained on thin films of the genuine hexagonal manganites (h-YMnO$_3$, h-ErMnO$_3$ or h-HoMnO$_3$) described above, thus suggesting that in h-TbMnO$_3$, the same tilting of MnO$_5$ bipyramids that lead to ferroelectric polarization in bulk hexagonal RMnO$_3$, is also operative in thin films. Naturally, the polarization of this h-TbMnO$_3$ is much larger than that of bulk o-TbMnO$_3$ (see below) where polarization does not originate from structural effects but from its magnetic structure. According to its hexagonal symmetry, the magnetic structure is expected to be also antiferromagnetic with ordering temperatures in the 60 -100 K. Unfortunately, but consistent with results in h-YMnO$_3$, h-ErMnO$_3$ and h-HoMnO$_3$, the magnetization measurements do not show sharp features at T$_N$ [40]. Similar comments hold for the magnetic response of h-DyMnO$_3$ [41]. Therefore, the Néel temperature of these "artificial" h-RMnO$_3$ films remains to be elucidated. Moreover, in both cases dramatic hysteresis ZFC-FC are observed in the temperature-dependent magnetization curves indicating the presence of a ferromagnetic component of unidentified origin.

### 4.2. Growth of o-RMnO$_3$ thin films

The epitaxial relationships in epitaxial heterostructures are primarily determined by the matching between the lattice parameters of the substrate and that of the material to be grown. The most commonly used single crystalline substrates for epitaxial growth of oxides have a cubic structure (i.e. SrTiO$_3$ (STO) or LaAlO$_3$ (LAO)) although orthorhombic substrates such as YAlO$_3$ (YAO) or NdGaO$_3$ (NGO) can also be employed. The mismatch is determined by the difference of lattice parameters of the substrate ($a_S$, $b_S$, $c_S$) and those of the material to be grown ($a_f$, $b_f$, $c_f$). Due to the orthorhombic nature of o-RMnO$_3$ oxides, the mismatch with cubic substrates is anisotropic and even of different sign, depending on the crystallographic direction considered. Therefore it can be expected that different crystallographic textures of films can be obtained by selecting the appropriate substrate and its crystallographic orientation. For instance, films of o-RMnO$_3$ on (001), (110) or (111) STO substrates may display distinct film textures. Similarly, films on equivalent crystalline planes of different cubic substrates, for instance (001)STO and (001)LAO can also lead to different textures due to differences of mismatch of o-RMnO$_3$ on both substrates. Similar considerations hold for orthorhombic substrates. In Table 1 we collect the out-of-plane textures achieved of o-RMnO$_3$ thin films grown on different substrates. Labels in the first row and in the first column indicate the symmetry of the equilibrium bulk phases of the some RMnO$_3$ oxides and of the substrates, respectively. Cell parameters of the metastable o-RMnO$_3$ (R = Ho-Y) phases, obtained from bulk samples prepared under high oxygen pressure, are included for completeness. Table 1, emphasizes that orthorhombic films of bulk hexagonal RMnO$_3$ have been obtained illustrating the power of epitaxial strain.

Table 1

| | | a (Å) | b (Å) | c (Å) | | **Bulk orthorhombic** | | **Bulk hexagonal** | | | | |
|---|---|---|---|---|---|---|---|---|---|---|---|---|
| | | | | | **R** | **Tb** | **Dy** | **Ho** | **Tm** | **Yb** | **Lu** | **Y** |
| | | | | | **a (Å)** | 5,3060 | 5,278 | 5,257 | 5,228 | 5,221 | 5,2 | 5,24 |
| | | | | | **b(Å)** | 5,8539 | 5,834 | 5,835 | 5,809 | 5,803 | 5,79 | 5,79 |
| | | | | | **c(Å)** | 7,4130 | 7,378 | 7,361 | 7,318 | 7,305 | 7,3 | 7,36 |
| **cubic** | STO(001) | 3.905 | 3.905 | 3.905 | [001] | [58-63] | [64] | [65-68] | | [69] | | [25, 70, 71] |
| | LAO(001) | 3,792 | 3,792 | 3,792 | [110] | | | | | | | |
| | STO(110) | 3.905 | 3.905 | 3.905 | [100] | | | [72] | [69, 73, 74] | [69, 75] | | |
| | LAO(110) | 3,792 | 3,792 | 3,792 | [010] | | | [76, 77] | | | [77] | |
| | STO(111) | 3.905 | 3.905 | 3.905 | [101] | | | | | [78] | | |
| **orthorhombic** | YAO(100) | 5,18 | 5,31 | 7,35 | [010] | [79, 80] | | | | | | |
| | YAO(010) | 5,18 | 5,31 | 7,35 | [010] | | | | | | | [81] |
| | YAO(110) | 5,18 | 5,31 | 7,35 | [110] | [82, 83] | | | | [84] | | |
| | NGO(110) | 5,431 | 5,499 | 7,71 | [110] | [85] | | | | | | |
| | NGO(001) | 5,431 | 5,499 | 7,71 | [001] | | | | | [75] | | |

It can be appreciated in Table 1 that in spite that distinct o-$RMnO_3$ oxides have different lattice parameters and thus slightly different mismatch with the different substrates, the observed thin film textures are independent on the precise $R^{3+}$ ion. The R-column, where we indicate the textures reported for the different o-$RMnO_3$ oxides, is a convenient guide for substrate selection when a particular thin film out-of-plane texture is required.

   As shown in Table 1, o-$RMnO_3$ films on (001)STO are *c*-textured, that is: the (001) planes of o-$RMnO_3$ grow parallel on the (001) planes of STO; however, these films face the intrinsic difficulty that twins should be present. Indeed, detailed studies of o-$TbMnO_3$ [59, 62], o-$YMnO_3$ [70] and o-$YbMnO_3$ [75] thin films have shown evidence of 4 or 2 variants of in-plane epitaxial relationships. For o-$TbMnO_3$ on (001)STO, the in-plane epitaxial relationship are $a$($TbMnO_3$)//[110]STO; $b$($TbMnO_3$)//[1-10]STO, that is, the o-$TbMnO_3$ unit cell (a similar situation holds for o-$YMnO_3$) is rotated by 45° with respect to the underlying square lattice of the substrate [62, 70].

   This is better seen in the sketches shown in **Figs. 8a-b**, where the epitaxial relationships and twin structure of o-$YMnO_3$ films on (001)STO and (110)STO are depicted. We note in **Fig. 8a** that there is an asymmetric in-plane strain acting on the film that shall depend on the particular o-$RMnO_3$ oxide. For instance, for $a$($TbMnO_3$)//[110]STO in (001)STO, the mismatch ($\varepsilon_{100} = [a_{TbMnO3} - a_{STO}\sqrt{2}]/a_{TbMnO3}$) is tensile (about -4 %) but compressive along $b$($TbMnO_3$)//[1-10]STO ($\varepsilon_{010} = [b_{TbMnO3} - a_{STO}\sqrt{2}]/a_{TbMnO3}$) (about +5.6 %); a similar situation holds for o-$YMnO_3$ (**Fig. 8a**) where the corresponding mismatches are -5% and +5.6 %. When growing o-$YMnO_3$ on (110)STO (**Fig. 8b**) it turns out that the *a*-texture is obtained $a$($TbMnO_3$)//[110]STO(110) and the in-plane c($TbMnO_3$) // [001]STO(110) and $b$($TbMnO_3$) // [1-10]STO(110). Again, the mismatch is anisotropic and of opposite sign along perpendicular in-plane directions: tensile (6.1%) along c($TbMnO_3$) and compressive along $b$($TbMnO_3$) [62].

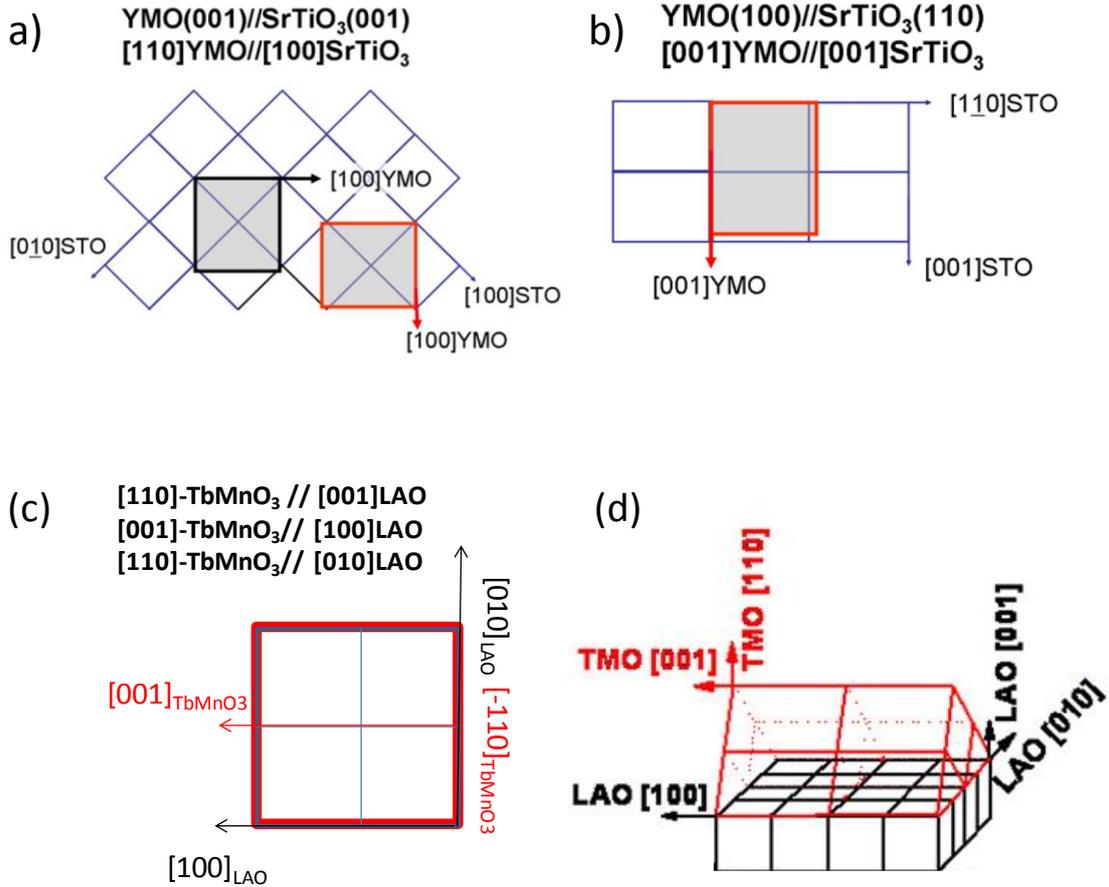

**Fig. 8** Epitaxial relationships for o-YMnO₃ films on: *a) (001)STO; b) (110)STO substrates. Adapted from [70] and (c,d) epitaxial relationships for o-TbMnO₃ films on LAO(001). Adapted from [86].*

When changing the cell dimensions of a cubic substrate, the mismatch with the growing film would differ and may induce a different texture. Indeed, for R= Tb the texture is [001] on (001)STO and [110] on (001)LAO [86] (see Table 1), thus implying that the (110) planes of o-RMnO₃ grow parallel on the (001) planes of LAO (Figs. **Figs. 8c,d**). The in-plane epitaxial relationship is found to be [001](TbMnO₃)//[100]LAO and [-110](TbMnO₃)//[010]LAO. Calculation of the mismatches for these directions signals a tensile strain ($\varepsilon_{001}$ = [$c_{TbMnO3}$ − $2a_{LAO}$]/$c_{TbMnO3}$) of about -2 %, for [001](TbMnO₃)//[100]LAO and compressive strain (≈ +4.9 %) along [110](TbMnO₃)//[010]LAO ($\varepsilon_{110}$ =[√ ($a^2_{TbMnO3}$+$b^2_{TbMnO3}$) − $b_{LAO}$]/[√ ($a^2_{TbMnO3}$+$b^2_{TbMnO3}$). A larger mismatch would exist if TbMnO₃ were to grow [001]-textured on (001)LAO.

On the other hand, using a higher index plane of cubic substrates (such as (110)STO) could have some advantages because the in-plane square symmetry is broken thus facilitating the growth of orthorhombic thin films. This approach has been used to obtain *a*-textured HoMnO₃ [72], TmMnO₃ [73, 74] and YMnO₃ [70, 71] films (**Fig. 8b**). A similar advantage is obtained by using orthorhombic substrates. This approach can be illustrated by taking as example the growth of TbMnO₃ on (001)YAO [79] or YMnO₃ on (010) YAO [81]. It can be appreciated in Table 1 that for TbMnO₃, the (a,c) cell-parameter (5.30 Å, 7.41 Å ) are very similar to the (b,c)-cell parameters of YAO (5.31Å, 7.35Å). Therefore, *b*-textured TbMnO₃ films are obtained [79]. Similarly, for YMnO₃ the *b*-axis of YMnO₃ is much larger than the others (*a, c*) (Table 1) and thus it would be preferable to grow the film with its *b*-axis out-of-plane to reduce mismatch and with the (a,c) axes in-plane. For this comparison is useful to considerer the pseudo-cubic cell parameters (*a*/√2, *b*/√2, *c*/2) of the film with those of the substrate.

Indeed, (010)YAO is very favorable substrate to obtain a $b$-textured [010]YMnO$_3$ film because the mismatches on (010)YAO are rather small (≈1.2% along [100] and < 0.1 % along [001] [81].

Selection of texture in o-RMnO$_3$ is of capital importance because multiferroicity is directly linked to the antiferromagnetic order and it can be expected that, if finite effects come to play a role, these should be more important for the out-of-plane direction. In bulk TbMnO$_3$, for instance, the antiferromagnetic cycloidal order propagates along the $b$-axis. If finite effects were to suppress the cycloidal order, it could be preferable exploring TbMnO$_3$ films with $b$-axis in-plane rather than for $b$ out-of-plane.

## 5. Magnetic and ferroelectric properties of o-RMnO$_3$ thin films

In this section we will revise in some detail the properties of some o-RMnO$_3$ thin films and we will compare with the properties of corresponding bulk materials. Before starting, it is worth to signal and comment some characteristics of the magnetization data of o-RMnO$_3$ thin films that may challenge data interpretation. Some of these issues also apply to the h-RMnO$_3$ discussed before. First, determining Néel temperature $T_N$ or other magnetic ordering temperature (i.e. the lock-in temperature ($T_{loc}$), see below) in thin films is somewhat challenging as the susceptibility of antiferromagnetic materials is small, and substrates and concomitant impurities may override genuine film features. In the case of RMnO$_3$ films where the $R^{3+}$ ions are magnetic, the presence of the large paramagnetic moment of the $R^{3+}$ ions, that largely increases the magnetization upon lowering the temperature, may rend challenging the observation of a cusp of susceptibility associated to any ordering of the much smaller moments of Mn$^{3+}$ ions. Moreover o-RMnO$_3$ films with $R^{3+}$ magnetic may display an anisotropic magnetic susceptibility extending far above the Néel temperature. This is mainly due to the quadrupolar moment of the $R^{3+}$ ions that give rise to strong anisotropy of the susceptibility in the paramagnetic state that may mask the features at $T_N$ differently depending on the magnetic field direction [87]. Therefore the observation of features at $T_N$ or $T_{loc}$ may depend on the anisotropic background provided by the $R^{3+}$ ions. Another consequence of the anisotropic paramagnetic $R^{3+}$ susceptibility is that the Curie-Weiss law give extrapolated paramagnetic temperature ($\theta_{CW}$) values that are not a direct measure of the strength of the magnetic interaction among the Mn$^{3+}$ ions [87]. Moreover, in an AF film, a genuine anisotropy of the susceptibility is expected below $T_N$: when applying the magnetic field along an axis perpendicular to the antiferromagnetic axis, i.e. the $a$-axis in o-RMnO$_3$ oxides, the susceptibility shall not produce a pronounced cusp at $T_N$ but should remain constant upon further lowering the temperature below $T_N$. Only when the field is applied at a direction along the antiferromagnetic axis (i.e. the $b$-axis in o-RMnO$_3$) the susceptibility should gradually vanish upon cooling. Therefore, in c-textured o-RMnO$_3$ films, the eventual presence of twins and their orientation with respect to the applied magnetic field should be considered when analyzing the susceptibility data. Second, some o-RMnO$_3$ films display a hysteretic response when performing ZFC-FC temperature-dependent magnetization M(T) measurements. This is a signature of the presence of uncompensated magnetic moments, giving rise to a ferromagnetic-like magnetization response that can also be seen in the magnetization $vs$ field M(H) loops. Third, at least in thin films grown by chemical vapor deposition, it has been shown that hausmannite (Mn$_3$O$_4$) precipitates can be formed during the growth [88]. Mn$_3$O$_4$ orders ferrimagnetically at about 41 K and thus it could contribute or mask any intrinsic effect in the o-RMnO$_3$ film. Note that the disturbing effects of Mn$_3$O$_4$, if present, should be less relevant in h-RMnO$_3$ films because of their higher Néel temperature.

We will first describe the properties of o-RMnO$_3$ films which bulk parents are E-type antiferromagents (R = Ho, Lu, Yb, Tm), next we will describe the results obtained from films whose corresponding bulk parents have a non-collinear (cycloidal) structure (R = Dy and Tb) and we will end with materials (o-YMnO$_3$), that being at the edge between both regions can be grown displaying either features of E-type or cycloidal magnetic ferroelectrics.

### 5.1. o-HoMnO$_3$ thin films

Early neutron scattering experiments on bulk o-HoMnO$_3$ showed that the material displays a transition from paramagnetic into an temperature dependent incommensurate collinear AF order at about 42 K that locks-in at T$_{loc}$ ≈ 26 K into an E-type magnetic structure. The magnetic order of Mn$^{3+}$ spins are stable up to about 2 K and a spiral order of Ho$^{3+}$ magnetic moments occurs at T$_{Ho}$ ≈ 6.5 K [89]. A finite polarization sets in at T$_{loc}$ with a large increase below T$_{Ho}$ reaching about 8.5 nC/cm$^2$ [90] in polycrystalline samples. Later measurements in single crystals gave ~ 0.24 μC/cm$^2$ [91].

As shown in Table 1, films of o-HoMnO$_3$ have been obtained with a-, b- and c-textures. The magnetization data of some films with a- , b- and c-textures displayed a cusp at about 42-44 K, followed by a smaller cusp at lower temperature (26 K - 35 K) respectively. The high temperature cusp is attributed to the Néel temperature (T$_N$ ≈ 42-44 K) where long range antiferromagnetic orders sets in. The low-temperature feature was claimed [65, 66, 92] to be more visible when the measuring magnetic H field is applied along the c-axis; however T. C. Han et al [72] found that the low temperature cusp was only visible for H//a. Although this feature could reflect the lock-in transition T$_{loc}$ observed in bulk, the observation of a susceptibility cusp for H//c is intriguing as the lock-in transition refers to spins in the ab-plane and it is unclear why it should produce visible effects on the c-axis susceptibility. The large magnetic anisotropy visible in magnetization data in some of the mentioned references, which extends to very high temperatures (up to about 150 K), although not discussed in any detail in the mentioned references, could be related to the paramagnetic anisotropy of the rare earths described above [87]. In all mentioned references (seen also in Wundelich [68] and [67]) the low-temperature magnetic susceptibility was found to display a pronounced hysteresis ZFC-FC, developing at temperatures roughly coinciding with T$_N$, thus leaving open the possibility that it arises from a genuine non-full compensation in at the antiferromagnetic region.

Ferroelectric polarization P(E) loops have been reported for c-textured o-HoMnO$_3$ films [65] and it has been claimed that polarization emerges at about 35 K, i.e. close (but well above) to T$_{loc}$, thus pointing to a connection between the E-type ordering and a possible ferroelectricity along c-axis. However, the observation of polarization along c-axis is not in agreement with predictions for an E-type magnetic ordering. The shape of the loops indicate that leakage may dominate the reported P(E) response and thus this result should be considered with caution. Permittivity ε(T) measurements on c-textured films [68, 92] and a-textured films [72], display an increase of permittivity at T ≈ T$_N$ and a broad maximum at T≈ 15 K which is depressed under large magnetic fields; although this ε(T) peak was assigned to a reordering of Ho$^{3+}$ magnetic moments [68, 92], its origin remains uncertain.

### 5.2. o-LuMnO$_3$ thin films

Neutron diffraction experiments showed that o-LuMnO$_3$ is AFM below 40 K and below 35 K it possesses an E-type magnetic structure [27]. Recent X-ray resonant scattering experiments revealed that within the E-phase, the spins are not purely collinear but there is a small spin canting [93] along the c-axis. The polarization of bulk samples has been reported to be of about 50 nC/cm$^2$ [94].

LuMnO$_3$ films with b- [77] and [110]-textures [84] have been reported. The magnetic susceptibility data of Tsai et al. [77] indicate an antiferromagnetic–like transition at T$_N$ ≈ 42 K superimposed to a dominant paramagnetic-like background which was observed to be strongly anisotropic. As Lu$^{3+}$ is not magnetic, the origin of this background and its anisotropy are unknown and were not discussed. In both cases, ferromagnetic-like responses were observed at low temperature. White et al. [84] performed neutron reflectometry and observed clear differences between spin-up and spin-down starting below 100 K and increasing upon cooling, signaling a ferromagnetic response. From this analysis the profile of magnetic moment μ(z) (z is the depth into the film) was obtained and it was concluded that a FM fraction of about 10 nm thick exists close to the film/substrate interface (with $\mu_{Mn} \approx 1.1\ \mu_B$) decreasing towards the film surface, that should coexist with a AFM fraction. In accordance, an exchange bias field was observed. The presence of the AFM dominant phase was confirmed by neutron diffraction experiments showing $T_N \approx 40$ K (**Fig. 9b**) and subsequent structural refinements indicated that the magnetic structure is not a simple E-type, as in bulk. Indeed the position of the magnetic reflections recorded at low temperature indicate an incommensurate magnetic order with $q_k$ = 0.482 (reciprocal lattice units) (**Fig. 9a**) instead of the expected $q_k$ = 0.5 for an E-type AF. Although the complete magnetic structure could not be refined, the results were argued to be compatible with a $bc$-cycloidal.

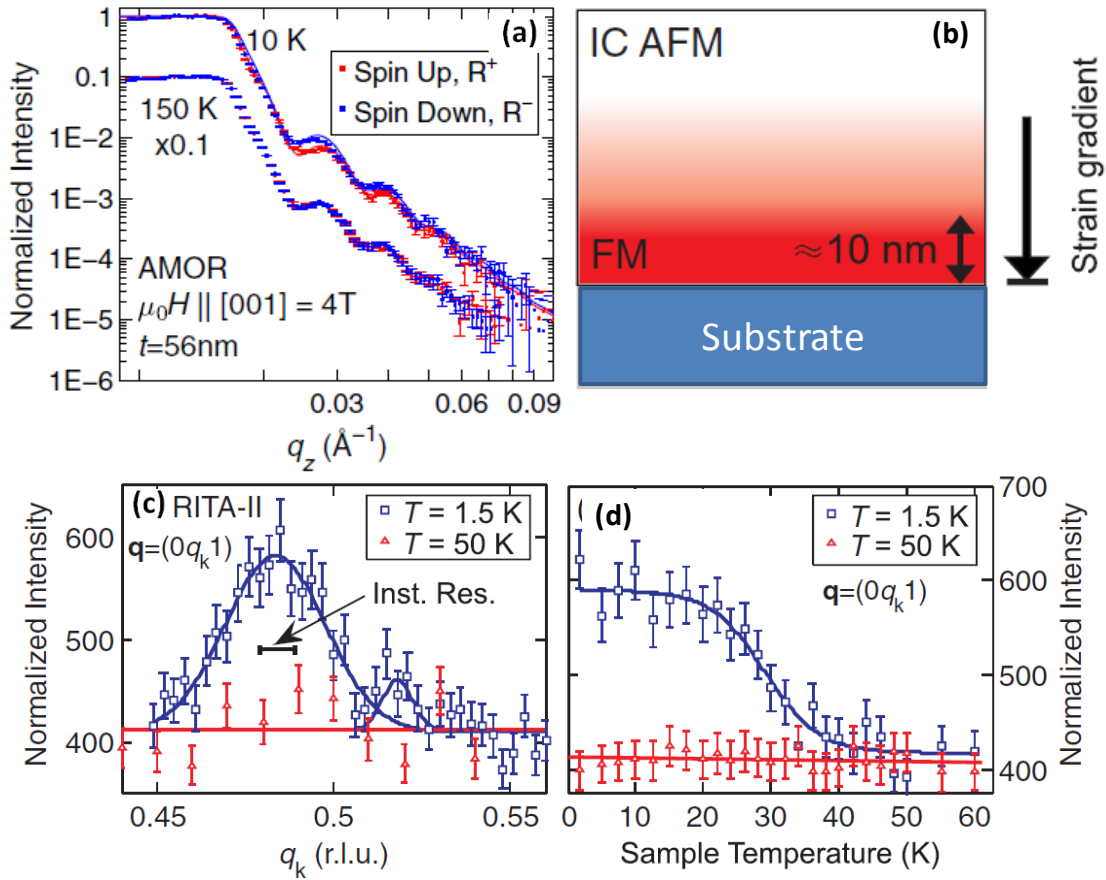

**Fig. 9.** (a) Intensities from PNR measurements as a function of $q_z$ at 10 and 150 K. The latter is multiplied by 0.1 for clarity. (b) Sketch to illustrate the magnetic response in the o-LuMnO$_3$ film. The FM layer located near the strained film-substrate interface evolves towards a likely cycloidal IC AFM order when moving down the strain gradient towards the film surface. (c) The (0q$_k$1) AFM peak in a o-LuMnO$_3$ thin film. A Gaussian fit of the main peak gives the peak center at $q_k \approx 0.482$. The solid bars indicate the instrumental resolution of RITA-II. (d) The temperature dependence of the (0q$_k$1) peak, with lines as guides for the eye. Adapted from [84].

Several messages emerged from this detailed study: a) the observation of a ferromagnetic distribution in the film, would not support that it emanates from Mn$_3$O$_4$ precipitates. The single crystalline nature of the [110] films [84], without twin domains, precludes a significant contribution of twin boundaries as a source of ferromagnetism as argued for $c$-textured TbMnO$_3$ films (see below). Instead, these results suggest that in presence of strain gradients, spin canting may arise due to unbalanced exchange interactions giving rise to a finite FM-like response; and b) the observation that magnetic ordering is not a pure E-type but display features compatible with a cycloidal one, may illustrate the sensitivity of these structures to epitaxial strain and may lead to either a coexistence of E-type and cycloidal order or a complete transformation of E-type order into to a cycloidal one, as theoretically predicted by Mochizuki et al [95]. A similar trend has been identified in o-YMnO$_3$ films (see below). No polarization data seem to have been reported.

### 5.3. o-YbMnO$_3$ thin films

Specific heat measurements on polycrystalline o-YbMnO$_3$ [96] showed transitions at about 40 K and 35 K indicative of the Néel temperature and the transition to E-type phase, respectively. Magnetic susceptibility only shows a tinny change of slope at about 43 K [97].

The magnetic properties of [101]-textured YbMnO$_3$ films have been reported by Rubi et al. [78]. It was found that the magnetic susceptibility, measured with in-plane field, was remarkably dependent on film thickness. Whereas a film of 15 nm displayed a cusp at about 40 K followed by a low temperature region where the ZFC and FC magnetization M(T) curves largely diverge, the susceptibility of a film of 38 nm was found to be paramagnetic-like without any other visible feature. Whereas the cusp in susceptibility observed in the thinnest film was attributed to the onset of the long range order AF order, the hysteretic response was attributed to a glassy behavior. The permittivity $\varepsilon$(T) displayed an upturn at about 42 K, the onset of AF order, and a broad peak with a maximum slope d$\varepsilon$(T)/dT at about 25 K which is close to a feature observed in specific heat of o-YbMnO$_3$ polycrystalline samples which were attributed to the onset of E-type ordering [96]. Similar features of $\varepsilon$(T) where observed in o-HoMnO$_3$ (described above), o-TmMnO$_3$ and o-YMnO$_3$ thin films (see below). To what extent the observation that the maximum slope of $\varepsilon$(T) coincides with the onset of E-phase in bulk materials is relevant or fortuitous remain to be addressed. Films with *a*- and *c*-textures [69] showed a similar susceptibility cusp at $T_N \approx 43$ K superimposed to a strong and anisotropic susceptibility background extending at least up to 140 K. Moreover, the susceptibility in the paramagnetic phase is found to be largely dependent on the texture of the film and strongly dependent on magnetic field. The observation of a field-dependent susceptibility in the paramagnetic state is at odds with expectation for a paramagnetic material and point to the presence of spurious ferromagnetic phases in these films. The presence of a strong hysteresis ZFC-FC observed at T < $T_N$ challenges a clear identification of further features occurring at lower temperature eventually signaling the setting the E-type order. No polarization data seems to have been reported.

### 5.4. o-TmMnO$_3$ thin films

In bulk o-TmMnO$_3$, antiferromagnetism sets in at about 42 K and evolves into an E-type AF order at $T_{loc} \approx 32$ K; polarization develops below $T_{loc}$ and, in polycrystalline samples, it amounts up to 0.15 µC/cm$^2$ [98].

o-TmMnO$_3$ films with *a*-axis perpendicular to film plane were first grown by Han et al [74]. The susceptibility data recorded with the field along the *a*-axis (normal to plane) showed a very tinny cusp at about 41 K and another one, only visible in the ZFC data, at about 32 K. The ZFC susceptibility remains weakly temperature-dependent up to about 5 K, as expected for an AFM material having the antiferromagnetic axis along *b*-axis. As these critical temperatures are very similar to those reported in bulk o-TmMnO$_3$ they were assigned to the equivalent magnetic transitions. The permittivity $\varepsilon$(T) displayed an upturn at about 42 K, i.e. coinciding with the onset of AFM order and a broad peak develops with a maximum about 25 K and a maximum slope of $\varepsilon$(T), occurring at 32 K that coincides with the onset of E-phase in bulk material. Temperature-dependent polarization measurements in films grown Nb:STO(110) revealed that spontaneous polarization start to grow at the lock-in temperature (32 K) reaching about 0.45 µC/cm$^2$ at 10 K. We strength that the polarization is recorded along the *a*-axis, as it is predicted to occur in E-type antiferromagnets. This value is larger than any other value reported in o-RMnO$_3$ films and it is larger than the value reported in bulk o-TmMnO$_3$.

### 5.5. o-DyMnO$_3$ thin films

Bulk o-DyMnO$_3$ displays a rather small ferroelectric polarization (around 0.2-0.3 μC/cm$^2$) along the c-axis (P$_c$) that can be switched to the a-axis (P$_a$) by application of a suitable magnetic field along b-axis [99]. This response is consistent with the existence of a bc-cycloidal antiferromagnetic order. Lu et al. [64], recently reported the growth of c-textured o-DyMnO$_3$ films on (001)Nb:SrTiO$_3$ substrates. The temperature dependence of the magnetization was measured with the field in the ab-plane (twinning precludes precise determination of the direction of the in-plane field with respect to the crystallographic axes) and along the c-axis. No signatures of the Néel temperature or other features related to magnetic ordering of Mn$^{3+}$ ions but only a bump at about T$_R$ ~ 9 K reminiscent of the ordering of Dy$^{3+}$ magnetic moments (**Fig. 10a**).

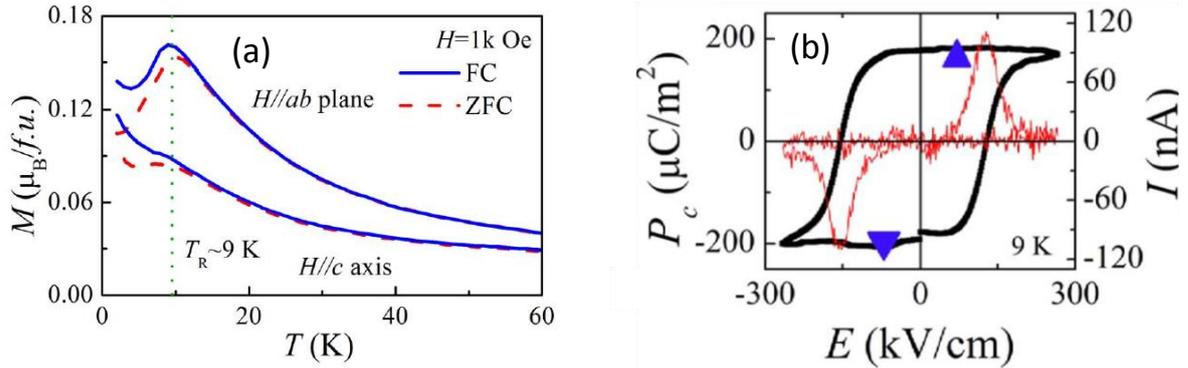

**Fig. 10** (a) *Temperature dependence of the magnetization measured in-plane and along the c-axis. (b) Ferroelectric hysteresis loop measured at 9 K without any magnetic field. Adapted from [64]*

The temperature dependence ε(T) reveals a subtle anomaly at about 40 K and a marked peak at about 21 K, both accompanied by hysteresis. Polarization measurements along c-axis indicate P$_c$ ≈ 0.02 μC/cm$^2$ (**Fig. 10b**). Both ε$_c$(T) and P$_c$(T) decrease below 9 K when Dy-Dy spin interactions dominate and spiral Mn$^{3+}$ magnetic moment order is suppressed. Importantly, when a magnetic field is applied along c-axis (H//c) there is no substantial effect on ε$_c$(T) and P$_c$(T) (**Fig. 11a**); in contrast, a significant variation of P$_C$ is observed if H//ab is applied at 2 K and the sample subsequently warmed up (**Fig. 11b**). In fact, P$_c$(H) first increase when increasing μ$_0$H up to 2 T and decreases by further increasing H (μ$_0$H > 5T). Similar effects were reported on o-DyMnO$_3$ singe crystals [99]. Whereas the decrease of P$_c$(μ$_0$H > 5 T) would be compatible with H-induced flopping of the cycloidal plane from bc to ab, the complex behavior observed at μ$_0$H < 5 K may reflect the competing effect of the Zeeman effect and the Dy-Dy and Dy-Mn interactions suggesting metamagnetic new spin phases induced by the field. Of relevance here is that the cycloidal order observed in o-DyMnO$_3$ singe crystals appears to be preserved in thin films. This is at difference with cycloidal order in other multiferroics, such as BiFeO$_3$ thin films where cycloids are suppressed either by confinement and/or strain [100].

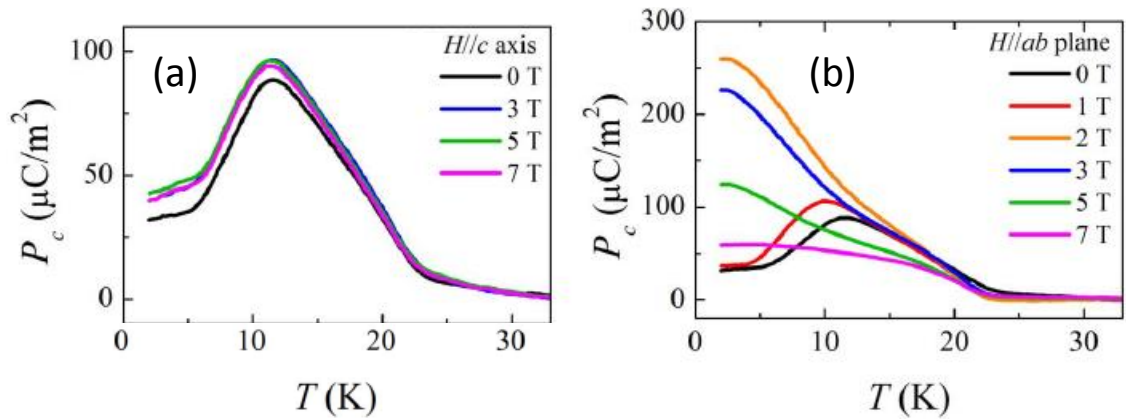

**Fig. 11** *Temperature dependence of the polarization Pc measured under various magnetic fields: a) H//c-axis and b) H//ab-plane. Adapted from [64]*

### 5.6.  o-TbMnO₃ thin films

At temperatures below $T_N \approx 42$ K, bulk o-TbMnO₃ has a sinusoidal incommensurate antiferromagnetic structure, with the propagating vector along the *b*-axis.  As the temperature decreases, the modulation vector varies until it locks at $T_{loc} \approx 27$ K, where the magnetic structure changes to a non-collinear cycloidal spin configuration and polarization appears along *c*-axis ($P_c$) [99]. Upon decreasing further the temperature, the $Tb^{3+}$ ions display a long range ordering at $T_{Tb} \approx 7$ K.  More recently, it has been shown that the magnetic structure of o-TbMnO₃ is more complex comprising a canted component of the magnetic moments, that orders itself with a sinusoidal or a cycloidal  structure, that develops both in the sinusoidal collinear phase and in the cycloidal one [101].

The magnetic properties of o-TbMnO₃ films were first reported on 2009 by Rubi [60] and Kirby et al [102]. The inverse magnetic susceptibility reported by Rubi et al [60] revealed large and negative extrapolated Curie-Weiss temperature ($\theta_{CW}$) ($\approx$ -150 K), a change of slope at about 40 K that was interpreted as signature of "ferromagnetic-like" transition (**Fig. 12a)**. and a splitting of the ZFC-FC magnetization curves below this temperature, indicating that a ferromagnetic component develops at low temperature. Kirby et al. [102] observed open magnetization *vs* field loops confirming the presence of a ferromagnetic component in o-TbMnO₃ films which was further asses by spin-polarized neutron reflectometry, performed at 6 K under $\mu_o H$ =0.55 T. From data fitting it was concluded that the magnetization profile was rather homogenous with some increase on magnetic moment at the bottom interface. The simplest conclusion from this observation is that the observed magnetism, in a sample expected to be antiferromagnetic, is not related to the presence of spurious phases (for instance, Mn₃O₄) as previously suggested. Quite intriguingly, the asymmetry of the spin-polarized neutron reflectometry was observed to persist up to 100 K [102].  Marti et al [62] reported a detailed study of the dependence of the magnetic properties of o-TbMnO₃ thins films on strain. The first remarkable difference with the previous studies is that the temperature $\theta_{CW}$ was found to be much lower ($\theta_{CW} \approx 0$), and in agreement with data from single crystals [103] and the extracted effective magnetic moment was fully consistent with that expected from the $Mn^{3+}$ and $Tb^{3+}$ ions (**Fig. 12b**). Anomalous and different $\theta_{CW}$ values may be due to uncertainties in background magnetization substraction.

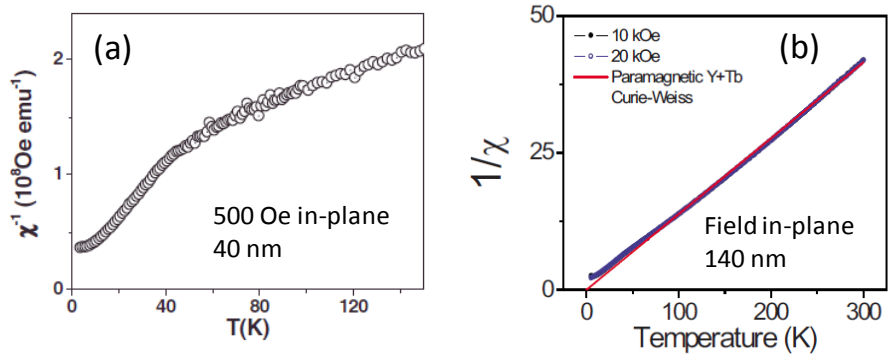

**Fig. 12** *Inverse susceptibility measured under after a FC process of o-TbMnO₃ films reported by (a) Rubi et al. [60] and (b) Marti et al. [62]*

Moreover the hysteresis ZFC-FC was found to be dependent on the film thickness and growth conditions (**Fig. 13a**) but, interestingly, it was observed that there was a continuous increase the ferromagnetic component that correlates with the unit cell volume compression modulated either by epitaxial strain or growth conditions (**Fig. 13b**).

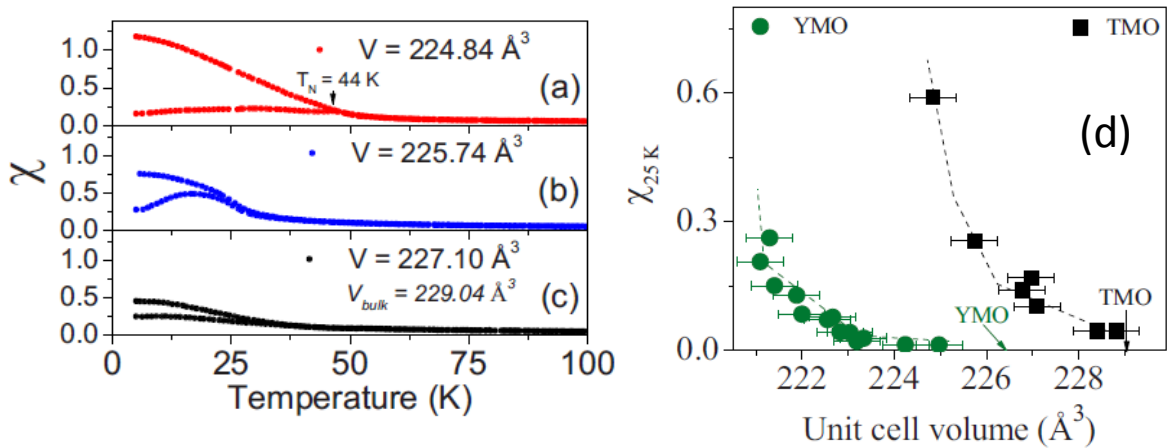

**Fig. 13** *(a, b, c) ZFC-FC magnetization curves of three o-TbMnO₃ films prepared under different conditions and labeled by the corresponding unit cell volume. (b) Dependence of the susceptibility (measured at 500 Oe) of a set o-TbMnO₃ films prepared either in different conditions or thickness, as a function of the unit cell volume. Adapted from Ref. [62]*

These observations are fully consistent with similar data collected in o-YMnO₃ thin films [104] (see below) suggesting that the ferromagnetic component arise from a strain-driven canted antiferromagnetic parent structure. Daumont et al [105] recently noticed a correlation between the ferromagnetism in *c*-textured o-TbMnO₃ films grown on (001)STO and the domain density (**Fig. 14**) and suggested that the domain walls formed as a result of twinning, rather than the domains, are the source of the net magnetization. Firmly setting this conclusion is challenged by the complexity of disentangling the relative contribution of domains and domains walls and a suitable crosscheck with magnetic anisotropy measurements [105]. Venkatesan et al. [63] used energy-loss electron spectroscopy to monitor changes in the valence state of $Mn^{3+}$ when approaching the interface with the substrates in *c*-textured *o*-TbMnO₃ on (001)STO and noticed a reduction in the nominal $Mn^{3+}$ oxidation state, but limited to the first 3-4 monolayers; although this observation could be of importance in the context of the interface polarity-mismatch problem, it cannot be of relevance to explain the dramatic ferromagnetic signal reported in these films [60]. Ultrafast optical spectroscopy has also been used to infer the presence of a ferromagnetic contribution in similar thin films [106].

Cui et al. [86] also reported magnetic data on films grown on (001)STO and (001)LAO. The resulting film texture is different (*c*- and [110], respectively) and both films displayed hysteretic ZFC-FC magnetization curves and rather unsatisfactory susceptibility data ($\theta_{CW} > 0$ and an unexplained existence of a field-dependent susceptibility up to very high temperature). Similar compressive strain state exists in the [110]-textured films grown on (110)YAO substrates reported by Hu et al [85]. Here, detailed X-ray diffraction experiments evidenced the co-existence of strained and relaxed sublayers in the films but unfortunately, evaluation and comparison of magnetic data was lacking.

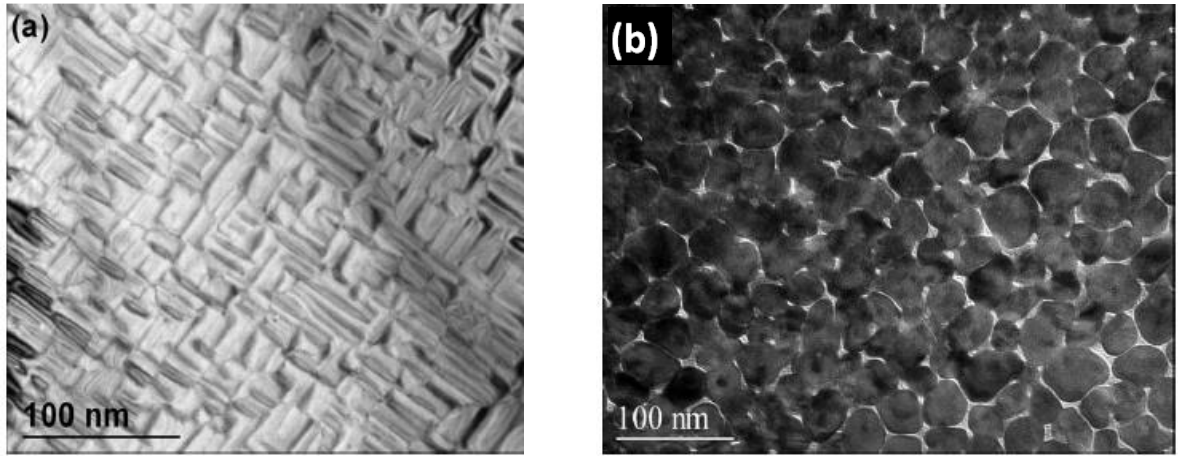

**Fig. 14** *Bright-field plane-view TEM image of a c-textured o-TbMnO$_3$ films on (001)STO. (a) 17 nm and (b) 140 nm. Adapted from [58]*

Results summarized above where collected on *c*-textured o-TbMnO$_3$ film grown on (001)STO, (001)LAO and (110)LAO and (110)YAO substrates which impose an overall compressive in-plane strain on the film.

Almost strain-free o-TbMnO$_3$ films can be obtained by growing films on (100)YAlO$_3$. Indeed, the basal plane dimensions (*b,c*) of the substrate (5.31 Å and 7.35 Å) match rather well with the (*a,c*) cell parameters of bulk o-TbMnO$_3$ (5.30 Å and 7.386 Å) thus anticipating a *b*-texture and a minor strain ($\approx 0.1$ % tensile and $\approx 0.4$% compressive along [100] and [001] directions of o-TbMnO$_3$). This approach has been followed by Glavic et al. [79] who reported neutron diffraction experiments showing that these films are antiferromagnetic below about 40 K in close agreement with bulk data. However, more recently, Glavic et al. [80] used x-ray resonant magnetic scattering (XRMS) techniques to show that these *strain-free* films develop a magnetic structure of the same kind that most recent magnetic structure refinement of bulk o-TbMnO$_3$ [101] and indicated the existence of cycloidal order below T$_{loc} \approx 27$ K, in films down to 6 nm thick (**Fig. 15a**). Notice that in these films the propagation direction (*b*-axis) is along the film normal, thus implying that finite effects (up to 6 nm) do not suppress formation of cycloidal order. Consistently, a second harmonic generation signal SHG (only allowed in non-centrosymmetric crystals) was observed to develop at temperatures below 27 K, reflecting the onset of polarization at exactly the same temperature where the cycloidal order sets in (**Fig. 15b**).

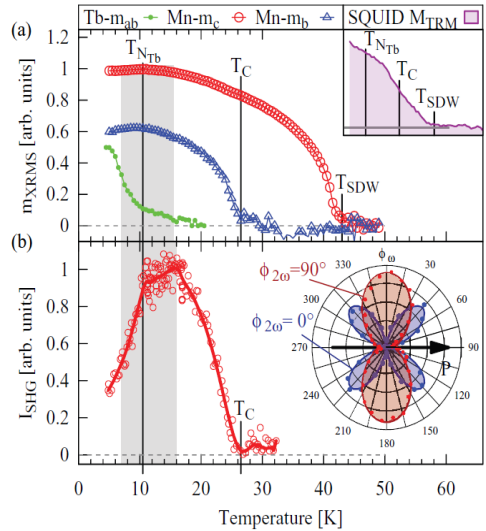

**Fig. 15** *Temperature dependence of the multiferroic order in a o-TbMnO₃ films. (a) XRMS and SQUID thermoremanent magnetization (inset) of a 11 nm thick film, and (b) by SHG of a 100 nm film. In (a), the b- and c- components of the Mn order and the ab component of the Tb order parameter are shown. (b) SGH intensity of the χ_caa component at 1.907 eV. The inset shows the polarization anisotropy of the SGH signal (at 11 K).* Adapted from [80].

The angular dependence of the SHG signal indicated a polarization along *c*-axis. Both the temperature where polarization develops and its direction are in full agreement with the polarization observed in the bulk. By using circularly polarized x-rays in XRMS, a different absorption for cycloids with clockwise and anticlockwise rotation should give rise to dichroism. Indeed, it was found that the dichroic signal disappeared above 27 K although magnetic scattering persisted up to T_N. Moreover by electric pooling of the crystal through T_loc it was demonstrated that the local sense of the polarization can be controlled and subsequent mapping of the dichroism allowed confirmed the switching of the sense of rotation of the cycloids. All these results constituted a solid demonstration that cycloidal order and the accompanying ferroelectric response and strong magnetoelectric coupling can be obtained in unstrained o-TbMnO₃ thin films. Still, these films displayed a finite thermoremanent magnetization which signals a ferromagnetic-like contribution emerging at T_N (see inset in Fig. 3a in [80]). As these films are non-strained, the origin of the observed magnetization should be related to either an intrinsic local canting or non-spin compensation due to defects in the film. In short, the origin of the finite magnetization in untwined strain-free remains an open question.

### 5.7. o-YMnO₃ thin films

Due to the small size of $Y^{3+}$ ions, bulk o-YMnO₃ develops an E-type magnetic structure below $T_N \approx 42$ [108] with a temperature-dependent incommensurate sinusoidal spin density that locks at about $T_{loc} \approx 28$ K and becomes ferroelectric at this temperature [90]. According to theory, polarization should develop along the *a*-axis and predicts a polarization of ~ 6 μC/cm² [34]. However, polarization measurements on polycrystalline o-YMnO₃ samples indicated that it is only of about $P \approx 0.5$ μC/cm² and samples may contain traces of cycloidal phase [94]. The possible coexistence of collinear (E-type) and cycloidal phase can be attributed to the close proximity of o-YMnO₃ to the boundary between collinear and cycloidal orders in the phase diagram.

Although stabilization of o-YMnO₃ films was reported already in 1998 by Salvador et al. [25], the magnetic properties of o-YMnO₃ films grown on (111)STO, STO(110) and (001)STO substrates, were first reported by Marti et al. [109]. The magnetic susceptibility of the *c*- textured film, measured with field in-plane, displayed a well defined peak at $T_N$ about 35 K, which is lower than bulk. As in *c*-textured o-YMnO$_3$ on (001)STO films at least two twin domains exist [70], the in-plane magnetic field cannot be applied selectively parallel nor perpendicular to the antiferromagnetic *b*-axis and thus the susceptibility is found to decrease below $T_N$ but not vanishing at the lowest temperature nor to remain constant below $T_N$ as expected in a untwined sample with the measuring field applied parallel or perpendicular to the antiferromagnetic axis. The antiferromagnetic character was also evidenced by growing films on a ferromagnetic SrRuO$_3$ bottom electrode and observing an exchange bias field vanishing at the Néel temperature, whose strength depends on the film texture and the concomitant domain wall density [70]. Hsieh et al [71] also reported on *a*-, *b*- and *c*-textured films, and observed that magnetic susceptibility of all films displayed a more or less pronounced cusp at about 44 K, signaling the Néel temperature. However, upon further cooling the susceptibility displayed an unexpected hysteresis ZFC-FC, very much depending on the orientation of the film (and substrate) signaling the presence of ferromagnetic contributions. Aiming to contribute to disentangle the origin of the anomalous ferromagnetic contribution, Marti el al [104] reported on the properties of *c*-textured films grown on (001)STO substrates, differing on film thickness and the oxygen pressure used during growth. They observed that the in-plane *a*-cell parameter is always relaxed but the *b*-parameter expand, the *c*-axis shrinks and the total unit cell volume expands when increasing thickness of O$_2$-pressure.  These lattice modifications largely impact the ferromagnetic-like component that increases when increasing in-plane strain or when the unit cell volume is reduced but reduces when recovering the unit cell volume by relaxing strain and growing films at higher pressure (**Fig. 16,** see also **Fig. 13b**). These observations strongly suggested that the strain-induced distortions of both *b* and *c* cell parameters, modifying the magnetic structure of o-YMnO$_3$ by inducing a non-compensated magnetic order, are at the origin of the observed weak ferromagnetic component. It was also found that this ferromagnetic contribution does not increases but decreases   when increasing film roughness and reducing the grain size, suggesting that its origin is not related to grain or domain boundaries but is the response of the domains themselves [110].

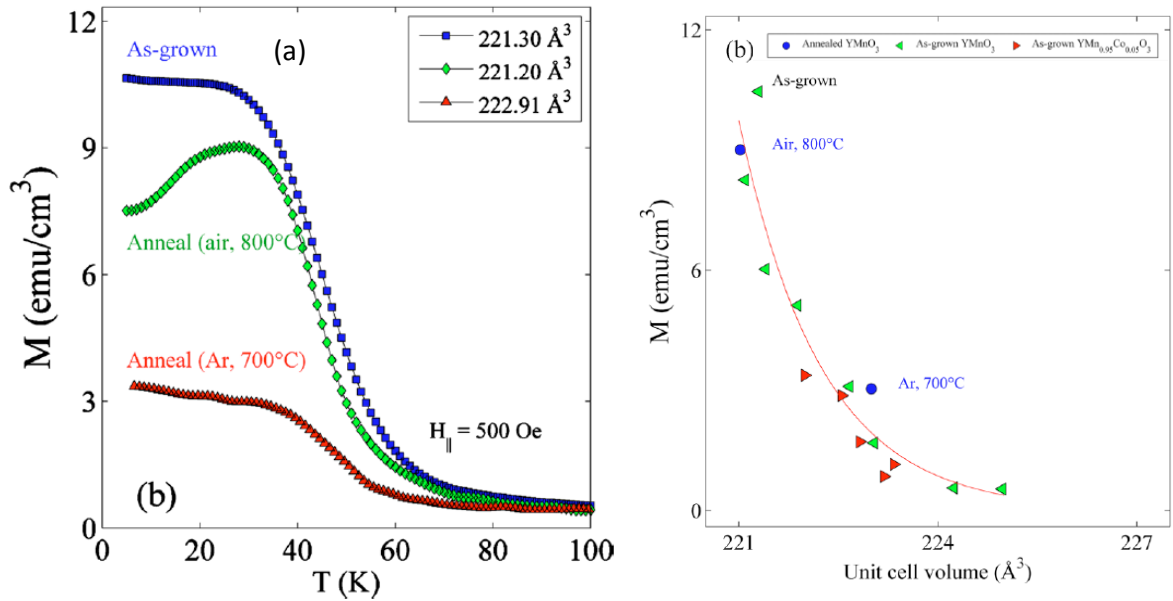

**Fig. 16** *(a) Magnetization (FC) of an o-YMnO$_3$ film: as-grown, and after air anneal (800°C) and argon anneal (700°C) processes. Labels indicate the unit cell volume values derived from X-ray diffraction. (b) Magnetization (at 25 K and 500 Oe; after a FC process) versus the unit cell volume for a number of as grown films (o-YMnO$_3$ (left-triangles) and o-YMn$_{0.95}$Co$_{0.05}$O$_3$ (right-triangles)) prepared either in*

*different conditions (temperature and oxygen pressure) and the annealed samples of (a) (circles). Adapted from [110].*

A deeper insight was obtained by exploring the magnetic anisotropy of thin films [110]. Using untwined *a*-textured films, it was shown that the hysteresis in the ZFC-FC data disappears if the magnetic field is applied in-plane along the *b*-direction but clearly shows up when the field is applied along any other direction (**Fig. 17**). It is thus concluded that the net magnetic moment is related to spin canting out of the *b*-axis. The canting angle was estimated to be of about 1.2° for the most strained films and decreasing when films are relaxed. These findings provided a rationale to describe the occurrence of ferromagnetic response observed in o-YMnO$_3$ films and, may be, in other orthorhombic manganite thin films. Early temperature dependent permittivity measurements $\varepsilon$(T) showed a broad peak developing below $T_N$ with a maximum at about 20 K, i.e. close to the lock-in transition of bulk o-YMnO$_3$ [111], which was found to be sensitive to magnetic fields and strain [112] and independent on frequency [113].

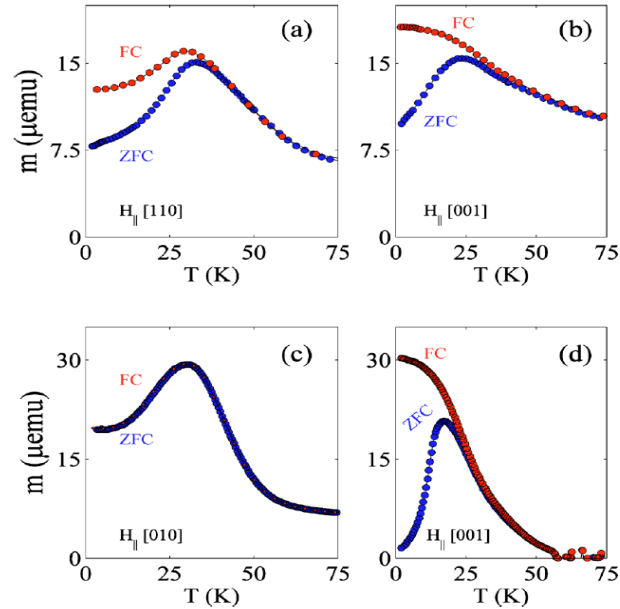

**Fig. 17** *ZFC-FC magnetic moment of twinned c-textured o-YMnO$_3$(001)/STO(001) films measured with H = 500 Oe applied (a) in-plane (H//[110]) and (b) out-of-plane ( H//[001]) directions. ZFC-FC magnetic moment of an untwined a-textured o-YMnO$_3$(100)/STO(110) films measured with H = 500 Oe, (a) in-plane (H//[010]) and (b) in-plane (H//[001]) directions. Data show that irreversibility (i.e. net magnetization) only occurs when the field is applied perpendicular to the b-direction. Adapted from [110].*

Polarization loops were reported on o-YMnO$_3$ films with c- and *a*-texture [4]. It turned out that a sizable polarization exists along the *c*-axis ($P_c \approx$ 90 nC/cm$^2$). Remarkably, a polarization develops along the *a*-axis ($P_a \approx$ 80 nC /cm$^2$) when a magnetic field is applied along the *c*-axis (H //c), whereas $P_c$ is reduced under H//ab [115, 116] These observations are characteristic signatures of a *bc*-cycloidal order switching to *ab*-cycloid as observed in Eu$_{1-x}$Y$_x$MnO$_3$ and (SmY)MnO$_3$ single crystals [94, 117, 118]. The persistence of domains of the minority cycloidal phase, after H-induced flop, with the majority one act as seeds for selecting a given chirality of the cycloidal winding upon field retreating and give rise to a remarkable chiral-memory effect [119]. Although similar results had been reported in single crystals of other o-RMnO$_3$ [117, 118] and related oxides [120], its observation in o-YMnO$_3$ thin films grown on STO(110) clearly assesses the existence of the cycloidal order in these compressive-strained films and denies a relevant contribution of E-type magnetic order on the observed ferroelectric response. Detailed analysis of the dielectric properties of differently strained films indicate a progressive suppression of ferroelectricity [115, 116, 121] that was interpreted as due to a progressive suppression of cycloidal order and a transformation of the films from E-type to A-type antiferromagnetic orders thorough a cycloidal region [121, 122].

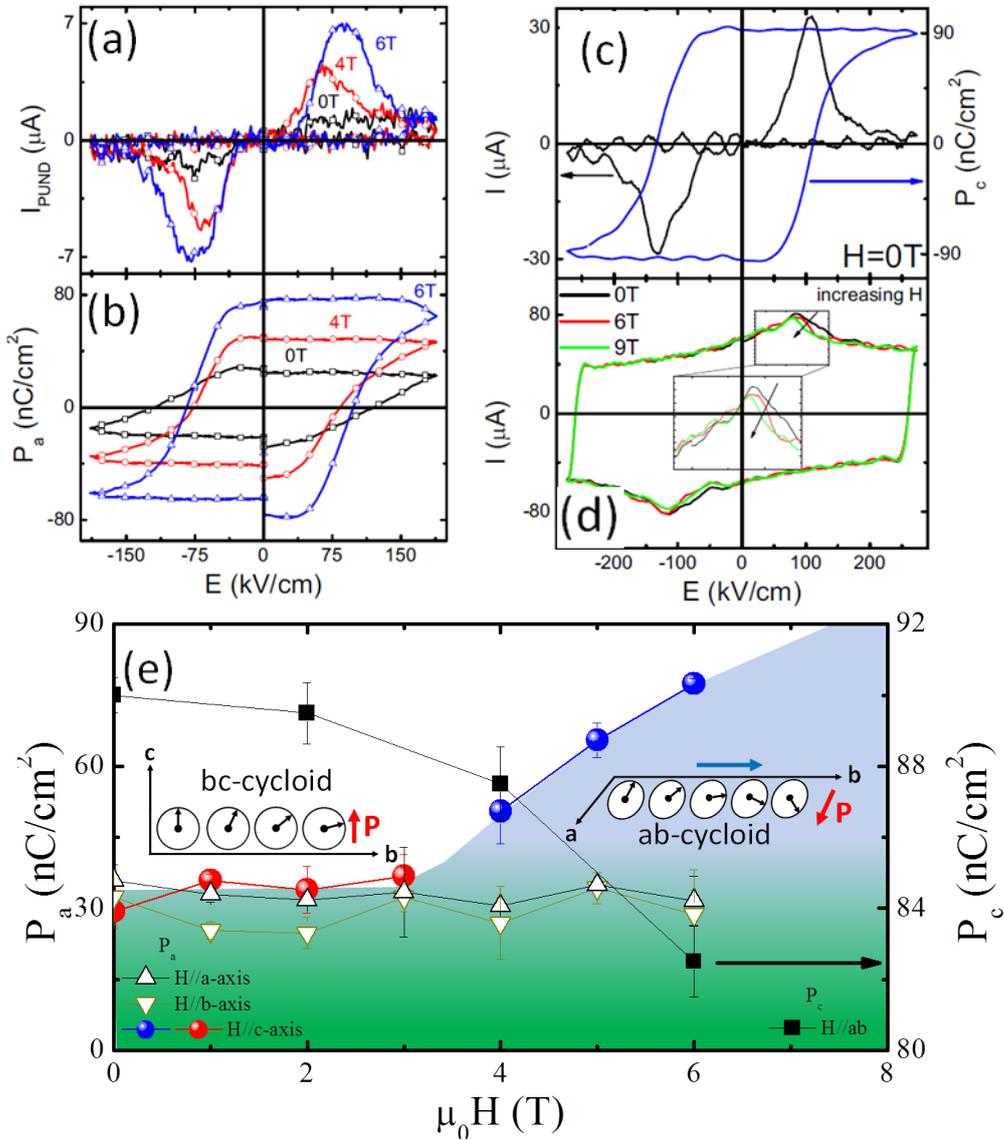

**Fig. 18** *(a) Current vs electric field $I_{PUND}$-E and (b) polarization $P_a$ (E) measured using PUND technique of an a-textured o-YMnO₃ film under various magnetic fields H//c. (c) current vs electric field I-E (left axis) and polarization $P_c$(E) (right axis) measured of a c-textured o-YMnO₃ film. (d) I-E curves recorded under various fields (H//c). (e) Phase diagram depicting the field-induced switch of polarization and the corresponding bc- to ab- flop of cycloidal plane. Adapted from [114].*

Recently, *b*-textured o-YMnO₃ have been grown on (010)YAO substrates with a somewhat smaller compressive strain as compared to films on STO [81]. Consistently with results described above, hysteretic ZFC-FC magnetization curves were measured for H perpendicular to the antiferromagetic axis but, in contrast, in zero magnetic field, polarization was observed to emerge along the *a*-axis $P_a$ (≈ 0.8 μC/cm²) but not along *c*-axis. The temperature dependence of $P_a$(T) and ε(T) revealed fine features at 40 K and 35 K, below $T_N$ (45 K). It was speculated that for 40 K <T< 45 K region, there is a coexistence of cycloidal (*bc*-type) and E-type and the observed step increase of polarization at lower temperature should correspond to the transition to a genuine E-type phase (**Fig. 19**).

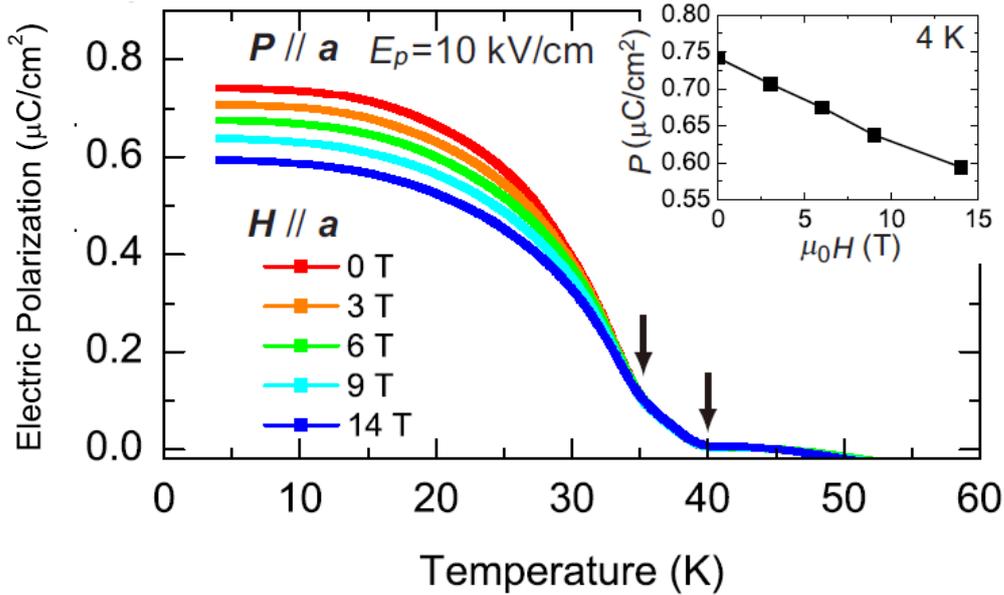

**Fig. 19** *Temperature dependence of electric polarization measured in various magnetic fields, after pooling in 10 kV/cm field. Arrows indicate the two successive ferroelectric transitions. The inset shows the magnetic-field dependence of the polarization at 4 K. Adapted from [81].*

Whereas the observation of a large $P_a$ is in agreement with predictions for E-type antiferromagnet, the assignment of the cycloidal fraction to *bc*-cycloids is difficult to rationalize and to convey with the observation of a finite $P_a$, it should probably read "*ab*-cycloid". It is clear that at the lowest temperature, presumably when the film has an E-type order, there is a significant coupling between the magnetic and the ferroelectric orders. It is intriguing however that in the temperature range 40 K-35 K where a cycloidal order is supposed to be dominant, there is no effect of the magnetic field on the polarization order as typically found. Subsequent experiments using resonant X-ray diffraction experiments [123], confirmed the coexistence of E-type and a cycloidal phase, although the technique could not discriminate between *ab* and *bc* cycloids and therefore, the nature of the cycloidal fraction in these films remains to be fully addressed. Although a canted spin component along c-axis was observed it appears to be fully compensated and it cannot explain the observed ferromagnetic-like signal.

### 6. Summary and conclusions

Progress on growth and characterization of magnetic and ferroelectric properties of h-RMnO$_3$ and o-RMnO$_3$ epitaxial films has been impressive. Epitaxial stabilization has allowed to obtain o- or h- films of genuine o- or h- materials but also to obtain metastable o- or h- films for virtually all lanthanides. Whereas first films of h-RMnO$_3$ films were grown in the 1960s with focus on polarization values and leakage for eventual applications on electronics, other relevant properties had not received much attention. For instance, the temperature and frequency dependence of the coercive fields or the ferroelectric domain structure of h-RMnO$_3$ thin films has been scarcely addressed. Similarly, systematic comparison of polarization in thin films with the corresponding bulk values, or as a function of strain remains to be done. On the other hand, it has been recently shown that h-RMnO$_3$ single crystals have a very intriguing 6-state vortex domain structure that emerges from the trimerization of Mn-ions network subsequent to the high temperature structural instabilities leading to ferroelectricity [124, 125]. These vortex structure have unexpected piezoelectric properties [126, 127] and the associated domain walls have been recently seen to form, at least in h-ErMnO$_3$, an array of alternating net magnetic moments pointing along *c*-axis, suggested to be due to uncompensated $Er^{3+}$ moments at domain walls [128]. Similar studies in thin films have not yet been reported. They could allow addressing important issues such as the existence of similar pattern domain vortex structures in thin films, and learn about how its density as a function of substrate induced strain or growth temperatures can be modified. The antiferromagnetic nature of h-$RMnO_3$ thin films, particularly when magnetic $R^{3+}$ ions are involved, is difficult to be monitored by standard magnetic susceptibility techniques and, although neutron diffraction has been successfully used in some few compounds, there is not clear information on how the Néel temperatures or the strength of the antiferromagnetic interactions in thin films are modified by the epitaxial strain, that itself is going to be dependent on the particular h-$RMnO_3$ under study. Disentangling the contribution of the rare-earth quadrupolar moment to the measured $\theta_{CW}$, as done in single crystals, would help to understand strain effects on the magnetic interactions. Last but not least, the persistent observation of ferromagnetic-like response emerging at $T_N$ remains to be fully understood. As indicated above, uncompensated $R^{3+}$ moments at domain walls could contribute to this ferromagnetic-like response, but this is restricted to c-axis magnetization and should not be present when the $R^{3+}$ ion is not magnetic. As overviewed here, this is against experimental observation and thus a more definitive answer is awaiting. The recent reports that canted antiferromagnetic structures and thus weak ferromagnetism, can be formed either in some multiferroic h-$RMnO_3$ or h-$RFeO_3$ oxides or even induced by magnetic field [129] may offer fresh views.

o-$RMnO_3$ films of virtually all lanthanides and other small radius cations (Y) have been obtained. The magnetic phase diagram of these compounds is extremely sensitive to changes of M-O-M bond angles and distances and the competing magnetic interaction among $Mn^{3+}$ ions, and to some extent the magnetic moments of $R^{3+}$ ions, therefore, not surprising, the progress on control and understating the properties of magnetic ferroelectric thin films has been slower. Probably, the most notable achievement is the observation that the E-type and cycloidal antiferromagnets can be obtained in thin films and convincing evidences of ferroelectricity in both cases have been obtained. Interestingly, cycloidal magnetic structures and the concomitant switchable ferroelectric polarization by suitable magnetic fields, have been obtained. Strain has been found to be a key factor to select a particular magnetic structure (E-type, ac- or bc- cycloidal) or even to suppress ferroelectricity by inducing an A-type antiferromagnetic ordering, although in some cases coexisting phases have been observed. Contribution of advanced tools, such as X-ray resonant magnetic scattering (XRMS), extended energy absorption fine structure (XAFS) or second harmonic generation microscopy has allowed to progress on some understanding on the magnetic structure of some thin films and beautifully demonstrate that cycloidals domains can be manipulated and used to store information in its winding state and associated polarization direction. Still fine details of the magnetic structure of thin films, for instance conclusive evidences and distinction from *ab* and *bc* cycloids and how strain affects these structures remain to be fully addressed. Magnetization data of o-$RMnO_3$ films, as much as the h-$RMnO_3$ films described above, display similar intriguing features, namely a ferromagnetic-like response emerging at $T_N$. Although it has been shown that these spurious signals scale with unit cell volume deformation thus pointing to a strain-induced canted magnetic structures, the debate with a possible contribution of domain walls continue and more work is required to settle this issue. As found in h-$RMnO_3$, multiferroic domains and domain walls can be a rich source of new physics. In cycloidal o-$RMnO_3$ magnetic ferroelectrics, the domain walls are genuinely different objects that bear no resemblance with ferroelectric domains walls or with magnetic domain walls due the intimate coupling of spin order and polarization. The properties and dynamics of these objects, received some attention in bulk materials, but it have been not yet explored in thin films.

**Acknowledgements**

This work was supported by the Spanish Government through MAT2011-29269-C03project and the Generalitat de Catalunya (2009 SGR 00376 and 2014 SGR 734 projects). The author is on duty to recognize the contributions of Ignasi Fina and Xavier Martí who did their PhD at ICMAB, on topics related to the present review. The author is also thankful to I. Fina for revising this manuscript.